\documentclass[11pt,a4paper]{article}
\usepackage{fancyheadings,lastpage}
\usepackage{pstricks,pst-node,pst-text,pst-3d}
\usepackage{color}
\usepackage{amssymb}
\usepackage{amsmath}
\usepackage{multirow}
\usepackage{multicol}
\usepackage{graphicx}
\usepackage{wrapfig}
\usepackage{hyperref}
\usepackage{amscd}
\usepackage{curves}
\usepackage{shadow}

\definecolor{cream}{rgb}{.97, .95, .88}
\definecolor{darkcream}{rgb}{1., .88, .5}
\definecolor{lightpink}{rgb}{.98, .88, .87}
\definecolor{lightwhite}{rgb}{1., .98, .95}
\definecolor{lightsalmon}{rgb}{1., .95, .90}
\definecolor{lightviolet}{rgb}{.99, .96, 1.}
\definecolor{lightgray}{rgb}{.96, .96, .96}
\definecolor{lgray}{rgb}{.75, .75, .75}
\definecolor{LemonChiffon}{rgb}{1., 0.98, 0.8}
\definecolor{lightolivegreen}{rgb}{.84, .89, .25}
\definecolor{lightgreen}{rgb}{.664, 1., .52}
\definecolor{llgreen}{rgb}{.900, .983, .960}
\definecolor{tristle}{rgb}{.792, .609, .698}
\definecolor{lighttristle}{rgb}{.892, .709, .798}
\definecolor{pink}{rgb}{.95, .35, .7}
\definecolor{magenta}{rgb}{1., 0, 1.}
\definecolor{violetl}{rgb}{0.820, 0.730, 0.969}
\definecolor{violet}{rgb}{.65, .30, .7}
\definecolor{darkolivegreen}{rgb}{.55, .65, .35}
\definecolor{maroon}{rgb}{.7, .26, .56}
\definecolor{mediumorchid}{rgb}{.8, .33, .83}
\definecolor{mediumorchidd}{rgb}{1., .33, .63}
\definecolor{darkgreen}{rgb}{0.1, .6, .13}
\definecolor{ddarkgreen}{rgb}{0.1, .45, .13}
\definecolor{lightyellow}{rgb}{1., 1., .82}
\definecolor{turquoise}{rgb}{.3, .70, .61}
\definecolor{turquoisel}{rgb}{.66, .94, .83}
\definecolor{darkturquoise}{rgb}{.21, .55, .50}
\definecolor{coral}{rgb}{1., .6, .21}
\definecolor{lightorange}{rgb}{1., .86, 0.69}
\definecolor{orangered}{rgb}{1., .5, 0.}
\definecolor{orange}{rgb}{1., .65, .1}
\definecolor{orangel}{rgb}{1., .85, .3}
\definecolor{darkorange}{rgb}{.875, .4, .204}
\definecolor{ddarkorange}{rgb}{.675, .318, .05}
\definecolor{bluesky}{rgb}{.48, .53, 1.}
\definecolor{gold}{rgb}{1., .85, 0.25}
\definecolor{goldd}{rgb}{.95, .75, 0.05}
\definecolor{darkgold}{rgb}{.85, .65, 0.05}
\definecolor{darkviolet}{rgb}{.54, .04, .84}
\definecolor{ddarkviolet}{rgb}{.382, .063, .657}
\definecolor{ddarkgreen}{rgb}{0.1, .45, .13}
\definecolor{lightblue}{rgb}{.30, .86, .89}
\definecolor{lblue}{rgb}{0.88, 0.90, 0.95}
\definecolor{LightBlue}{rgb}{0.68, 0.85, 0.9}
\definecolor{darkblue}{rgb}{.105, .308, .707}
\definecolor{lightmaroon}{rgb}{0.85, 0.38, 0.58}
\definecolor{darkmaroon}{rgb}{0.604, 0.169, 0.451}
\definecolor{darkpink}{rgb}{0.879, 0.020, 0.766}
\definecolor{ddarkpink}{rgb}{0.738, 0.195, 0.406}
\definecolor{grey}{rgb}{0.717, 0.717, 0.717}
\definecolor{brown}{rgb}{0.640, 0.323, 0.182}
\definecolor{darkbrown}{rgb}{0.34, 0.25, 0.05}
\definecolor{orangebrown}{rgb}{0.433, 0.262, 0.06}
\definecolor{pinkl}{rgb}{1., 0.788, 0.918}
\definecolor{salmon}{rgb}{1., 0.66, 0.5}
\definecolor{lightbrown}{rgb}{0.703, 0.508, 0.121}

\def\etal{{\it et al.}}

\def\Journal#1#2#3#4{{#1} {\bf #2}, (#3) #4}

\def\AA{\em A.\& A.}

\def\AIP{\em AIP Conf.Proc.}

\def\APH{\em Annals Phys.}

\def\APJ{\em ApJ.}
\def\APJL{\em ApJ.Lett.}

\def\APP{\em Astropart. Phys.}

\def\ASB{\em Ann. Soc. Sci. Brux. A}
\def\ASP{\em ASP Conf.Ser.}

\def\AST{\em Astron. J.}

\def\CQG{\em Class.Quant.Grav.}

\def\ENT{\em Entropy}
\def\EPL{\em Europhys. Lett.}
\def\FOP{\em Found. Phys.}

\def\GRG{\em Gen. Rel. Grav}
\def\IET{\em IEEE Trans. Comm.}
\def\IMA{{\em Int. J. Mod. Phys.} A}
\def\IMD{{\em Int. J. Mod. Phys.} D}

\def\ITP{\em Int. J. Theor. Phys.}
\def\JCA{\em J. Cosmol. Astrop. Phys.}
\def\JHE{\em J. High Ener. Phys.}

\def\JPC{\em J. Phys.: Conf. Series}

\def\JPL{\em JETPhys. Lett.}
\def\JPM{\em J. Phys.: Condens. Matter}

\def\MPL{{\em Mod. Phys. Lett.} A}
\def\MRA{\em MNRAS}

\def\NAT{\em Nature}

\def\NJP{\em New J. Phys.}

\def\NPB{{\em Nucl. Phys.} B}

\def\PLB{{\em Phys. Lett.} B}
\def\PNS{\em Pub. Nation. Acad. Sci.}

\def\PRB{{\em Phys. Rev.} B}

\def\PRD{{\em Phys. Rev.} D}
\def\PRL{\em Phys. Rev. Lett.}
\def\PRV{\em Phys. Rev.}
\def\PRE{\em Phys. Rep.}

\def\RMP{\em Rep. Mod. Phys.}

\def\SCI{\em Science}

\def\SSR{\em Space Sci. Rev.}

\def\be{\begin{equation}}
\def\ee{\end{equation}}
\def\bea{\begin{eqnarray}}
\def\eea{\end{eqnarray}}
\def\bes{\begin{equation*}}
\def\ees{\end{equation*}}
\def\beas{\begin{eqnarray*}}
\def\eeas{\end{eqnarray*}}
\def\phiq{{\phi}_q}
\def\phix{{\phi}_x}

\def\mg{\mathsf g}
\def\cx{\mathtt X}
\def\ca{\mathtt A}
\def\um{\mathcal U}

\def\ym{\mathcal Y}

\begin{document}
\vspace{2cm}
\begin{center}
{\Large \bf Classical, quantum, and phenomenological aspects of dark energy models}\\
\bigskip
Houri Ziaeepour\\
\bigskip
\begin{tabular}{c}
\multirow{2}{12cm}{\it Institut UTINAM, CNRS UMR 6213, Observatoire de Besan\c{c}on, 41 bis ave. 
de l'Observatoire, BP 1615, 25010 Besan\c{c}on, France} \\
\\
\\
{\tt houriziaeepour@gmail.com} \\
\end{tabular}
\end{center}
\vspace{2cm}
\begin{abstract}
The origin of accelerating expansion of the Universe is one the biggest conundrum of fundamental 
physics. In this paper we review vacuum energy issues as the origin of accelerating expansion 
- generally called dark energy - and give an overview of alternatives, which a large 
number of them can be classified as {\it interacting scalar field} models. We review properties of 
these models both as classical field and as quantum condensates in the framework of non-equilibrium 
quantum field theory. Finally, we review phenomenology of models with the goal of discriminating 
between them. 
\end{abstract}

\pagebreak
\tableofcontents
\pagebreak
\pagestyle{fancyplain}
\section{Introduction} \label{sec:deqft-intro}
The discovery of dark energy is one the most incredible and fascinating stories in the history of 
science. In 1917 A. Einstein added an arbitrary constant $\Lambda$ called {{\it the 
Cosmological Constant} to his equation to obtain a static solution for a homogeneous 
universe~\cite{einsteinlambda}. In 1924 A. Friedman~\cite{friedmanprove} proved that the static 
solutions of the Einstein equation are unstable and even the slightest fluctuation of matter 
density leads to a collapse or an eternal expansion. The same result was obtained by 
G. Lema\^itre~\cite{expanfrench} who explained the newly discovered redshift of distant galaxies 
by V. Slipher and E. Hubble~\cite{firstredshift}, as the expansion of the Universe in agreement with 
the prediction of Friedman and his own calculation of the expansion rate - the Hubble constant $H_0$. 
In addition, in 1917 W. De-Sitter~\cite{desittercosmos} showed that even in absence of matter when 
$\Lambda \neq 0$, the Universe expands if $\Lambda > 0$ or collapses if $\Lambda < 0$. This is in 
contradiction with Einstein gravity theory which associates the curvature of spacetime to matter. 
Apparently in early 1920's Einstein regretted the addition of the Cosmological Constant to his famous 
equation. Nonetheless, in a letter to him, Lema\^itre considered the idea as genius and interpreted 
it as {\it the energy density of the vacuum}~\cite{lemaitrevac}. Since the 
introduction of this interpretation, we are struggling to understand what is the meaning of the 
counter-intuitive claim of a vacuum that carries energy, and what its value may be.

For roughly 70 years, depending on the taste of authors, a cosmological constant was added or removed 
from Einstein equation. For instance, in the introduction of the famous book {\it Gravitation} by 
C.W. Misner, K.S. Thorne and J.A. Wheeler written in early 1970's, the authors compare the Cosmological 
Constant with the Pandora Box and say that despite its futility, people continue to discuss it. 
They mainly neglect the Cosmological Constant through their book except in a few places. Therefore, 
it was a great surprise when in the middle of 1990's precise measurements of cosmological parameters 
from anisotropies of the Cosmic Microwave Background (CMB)~\cite{cobe,saskatoon} and Large Scale 
Structures (LSS) of the Universe~\cite{apmsurvey}, and direct measurements of $H_0$ using 
Cepheids variable stars~\cite{h0cepheid} showed that the Universe is flat but there is not enough 
matter to explain the expansion rate which is too large for a flat matter dominated Universe. 
Such a model leads to a universe younger than some of the old globular clusters in the halo of the 
Milky-Way and old elliptical galaxies~\cite{starage}. In 1998-1999 
observations of supernovae type Ia showed that the expansion of the Universe is accelerating. 
According to the Einstein general relativity such a state is consistent only if the average energy 
density of the Universe is dominated by a cosmological constant or something that behaves very 
similar to it, at least since redshift $z \sim 0.5$ i.e. about half of the age of the Universe when 
its energy density became dominant. Therefore, a quantity added by hand which failed to provide its 
initial aim turned to be the dominant constituent of the Universe today !

Even before the confirmation of the presence of a cosmological constant or something that very 
closely imitates it, people had considered the issues that a cosmological constant 
imposes on our understanding of fundamental physics and cosmology~\cite{weinbergde}. They will be  
discussed in some details in the next section. Here, we briefly review alternative models which 
are suggested as the origin of the observed accelerating expansion of the Universe. They are 
summarized in Fig. \ref{fig:demodel}.
\begin{figure}
\begin{center}
\includegraphics[width=12cm]{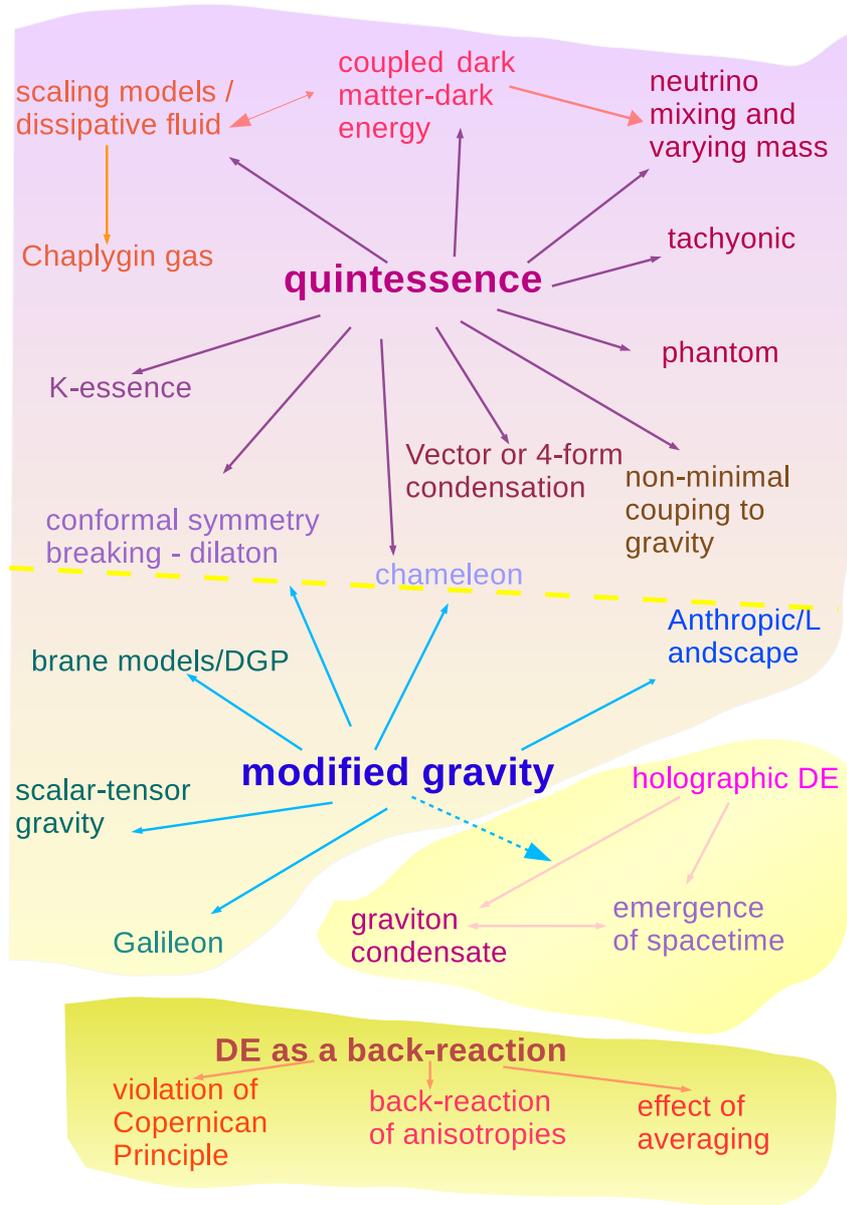}
\caption{\sf \small Alternative models of dark energy. The two main categories are quintessence and 
modified gravity. Related models are marked by arrows. Their differences are in the details of 
interaction, potential, symmetries, etc. Some models 
can be considered as alternative formulations rather than fundamentally different models. For 
instance, a dissipative fluid is similar to an interacting quintessence formulated as a fluid 
rather than a field. The dash line between quintessence and modified gravity models indicates 
the fact that many of specific cases of both models are based on a scalar field and distinction 
between the two needs more criteria, see also Sec. \ref{sec:deqft-int}. Note that models in this 
categories can be treated either classically or as quantum models. There are also two smaller 
separate groups: holographic models that use holography principle in the context of semi-classical 
cosmologies, and a last group of models in which dark energy is assumed to be an induced effect 
due to the averaging of fluctuations or negligence of higher order anisotropies. The color of fonts 
schematically presents closeness of models. \label{fig:demodel}}
\end{center}
\end{figure}
Quintessence models are based on a scalar field which in the original proposal only interacts with 
itself. For polynomial potentials of negative order or exponential potentials with negative 
exponent a class of solutions called tracking exists, meaning that at late times they approach to a 
small but nonzero value of the potential. The equation state of dark energy parametrized as 
$P = w\rho$ where 
$P$ and $\rho$ are pressure and density, respectively. In generic form of quintessence models 
$w \geqslant -1$. However, estimation of cosmological parameters from combination of supernovae, 
LSS and CMB data seems to prefer $w \lesssim -1$. For this reason, various extensions of pure 
quintessence models are proposed. They have either a non-standard kinematic term in their Lagrangian 
or interaction with other components, notably with dark matter or neutrinos.

Including the Cosmological Constant term in the geometry side of the Einstein equation conceptually 
makes it part of gravity. Therefore, an alternative to a cosmological constant may be a modification 
of the Einstein gravity at cosmological scales. This idea is favored by many authors who consider 
both a cosmological constant and a quintessence model as superficial. It is also somehow supported 
by history. Einstein gravity was introduced after the discovery of deviation from Newtonian 
gravity. Another reason is the fact that curvature dependence in Einstein equation is minimal and 
higher order models such as Gauss-Bonnet and scalar-tensor models have been suggested decades before 
the discovery of dark energy. Additionally, models inspired by string and brane theories have 
revived the idea of modification of the Einstein gravity. Notably, in  brane models these 
modifications can be at cosmological scales. The most famous model in this category is DGP that 
associates dark energy to induced terms from boundary conditions imposed on the 4D spacetime of the 
visible brane by 5D bulk in which gravity but not other fields can propagate.

The two other sets of models presented in Fig. \ref{fig:demodel} are newer and have less 
supporters, and consequently less studied. Holographic models are based on the Bekenstein limit 
on the maximum amount of entropy in a closed volume and holographic principle. The large entropy of 
the Universe seems to violate this conjunction. To solve this apparent contradiction~\cite{quinholog} 
holographic models of dark energy assume a relation between UV and IR cutoffs in the determination 
of vacuum energy density. In this case the vacuum energy density becomes a constant comparable to 
the observed dark energy. This apparently simple solution has several problems, notably the relation 
between scales is arbitrary and it cannot provide $w \lesssim -1$ because the entropy or temperature 
would be negative.

As for the effect of anisotropies, the claim is that we live inside an anisotropies universe and 
by averaging determine the homogeneous cosmological quantities such as the present value of Hubble 
constant $H_0$ and the fraction of matter density $\Omega_m$ . This can induce an error in the 
estimation of cosmological quantities because we do not access on the totality of the Universe at the 
same time. The correctness of the argument is evident, but can this error 
be enough large to induce a large {\it effective dark energy} which at present has a density close 
2.5 times the matter that produced it - according to this suggestion ? Some supporters of such an 
explanation think superhorizon modes, i.e. modes that after being pushed outside the horizon by 
inflation, are not yet entered inside can cause such a large effects. Other supporters believe 
that the effect of LSS, i.e. modes which are already inside horizon is dominant and explains the 
apparent observation of dark energy. In Sec. \ref{sec:backreact} a short commentary letter~\cite{p45} 
about this model. Finally, the last model in this category suggests that we leave 
in a locally under-dense region of the Universe.

In the following sections we review vacuum problem ~\cite{p2} and dark energy models studied 
in~\cite{p29,p25,p20,p50,p8}.

\section{Vacuum energy} \label{sec:deqft-vac}
\subsection{Introduction}
If dark energy is {\it the energy density of vacuum} as Lema\^itre suggested, we must put 
forward a precise definition for what we call {\it vacuum}. The dictionary meaning of this word is 
{\it emptiness}. In quantum field theory a vacuum state is more subtle. For instance, the minimum 
of the potential - the ground state - of a field is also called a {\it vacuum}. For a free field 
the vacuum $|0\rangle$ is defined as:
\be
a_\alpha~|0\rangle = 0, \quad \forall \alpha \label {vacdef}
\ee
where $a_\alpha$ is the annihilation operator of mode (state) $\alpha$ which presents the set of all 
quantum numbers of the field. Considering for simplicity $\alpha = k$ (the momentum), the energy 
density is the $T^{00}$ component of energy-momentum tensor $T^{\mu\nu}$ and in a locally flat space 
can be related to modes as the following:
\bea
&& \langle 0|\hat{T}^{00}|0\rangle = \langle 0|\frac{1}{(2\pi)^3} \int d^3k~u_ku^*_k 
\omega_k (a_k a_k^\dagger + a_k^\dagger a_k)|0\rangle = \frac{1}{(2\pi)^3} \int d^3k~
\omega_k \rightarrow \infty, \quad \omega_k = \sqrt {\vec{k}^2 + m^2} \label{t00minkovski} \\
&& [a_k,a_{k'}^\dagger] = \delta_{kk'} \label {commutrel}
\eea
Here $u_k$ and $u^*_k$ are solutions of the field equation for the set of parameters 
${\{\alpha\}} = k$. For fermionic fields the commutation relation in (\ref{commutrel}) is replaced by 
an anticommutation. Because in Minkowski space there is a Killing vector for the whole spacetime, 
conjugate functions $u_k$ and $u^*_k$ are independent solutions of the field equation and there is 
no ambiguity in the definition of particles and anti-particles. Thus, a {\it natural (adiabatic)
\footnote{A vacuum is called adiabatic if no particle or only particles with $k \rightarrow 0$ are 
created during the evolution of spacetime~\cite{qftcurve}.}} definition for vacuum exists. By 
contrast, in expanding spaces such as FLRW and De Sitter, there is no unique vacuum~\cite{desittervac}. 
Nonetheless, it can be shown that these apparently different vacua correspond to adiabatic vacuum in 
frames moving with respect to each other - in general with varying velocities~\cite{qftcurve}. They 
are related to each others by a Bogoliubov transformation.

The singularity in (\ref{t00minkovski}) is due to the ambiguity of operators $\phi^2$ and 
$(\partial^0 \phi)^2$ in $\hat{T}^{00} = 1/2 \partial^0 \phi \partial^0 \phi + 1/2 m^2 \phi^2$ when 
$\phi$ is a quantum field~\cite{qftcurve}. An operator ordering: 
\be
\phi^2(x) \rightarrow 
:\phi^2 (x):~\equiv~\lim_{y \rightarrow x}~\biggl \{ \phi (x) \phi (y) - \langle 0|\phi (x) \phi (y)
|0\rangle \biggr \} \label{opord}
\ee
(and the same for the derivative term) or another regularization brings the vacuum energy to zero. 
However, it is considered that in curved spacetimes where in contrast to flat spaces the zero-point 
of energy is not arbitrary, the application of regularization techniques to energy-momentum tensor 
is ad hoc~\cite{descalar}. 

In a recent work~\cite{p2} new interpretations are proposed for the ambiguity of the definition 
of vacuum and the singularity described above, and suggested a new definition which is frame 
independent.

\begin{wrapfigure}{r}{0.3\textwidth}
\vspace{-2.5cm}
\begin{center}
\includegraphics[width=0.3\textwidth]{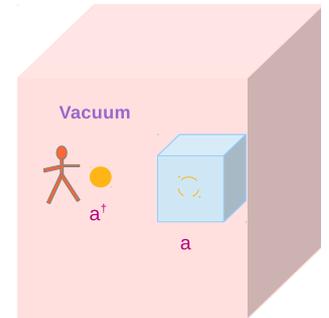} % width is the fraction of text width
\end{center}
\vspace{-2cm}
\caption{\sf \small A schematic description of the confinement of a created particle before annihilating 
it. The dot (orange) presents the particle created by operator $a^{\dagger}$ and the dash circle presents 
its annihilation by $a$ after its confinement in the blue volume that induces a Casimir effect. 
\label{fig:casimir}}
\end{wrapfigure}

\subsection{New interpretations}
To better understand the physical meaning of regularization consider the operator $a_k a_k^\dagger + 
a_k^\dagger a_k$ in (\ref{t00minkovski}). The second term is the number operator $\hat {N}_k$ and by 
definition $\hat {N}_k|0\rangle = 0 |0\rangle ~\forall~k$. Therefore its application does 
not change the state. Moreover, operationally it is well defined. the operation can be considered 
as creation of a particle in mode $k$ and its immediate annihilation. In a classical view these two 
operations are opposite to each other and leave the space unchanged if the delay between these 
operations is negligible. The first term is $\hat {N}_k +1$ and its constant part leaves a 
{\it remnant} energy which is the origin of the singularity in (\ref{t00minkovski}). 
This can be interpreted as an error aroused from using the classical expression of $T^{\mu\nu}$, which 
as explained above, in a quantum context is not well defined. In this case the operator ordering or 
other regularization schemes seem legitimate irrespective of the geometry of spacetime.

Regarding the operational description of energy measurement, the application of 
$a_k^\dagger$ creates one particle with momentum $\vec{k}$. If we exactly know the momentum of the 
particle, all information about its position are lost. Therefore, an observer who wants to 
apply the annihilation operator must first somehow localize the particle, otherwise the 
probability of annihilation becomes negligibly small. Such an operation would not be possible 
without breaking the translation symmetry of the spacetime, for instance by imposing boundaries 
at a distance $L \sim 1/k$ which induces a Casimir energy proportional to $1/L \sim k$, and 
becomes infinite for $k \longrightarrow \infty$. This shows that in contrast to some 
suggestions~\cite{weinbergde}, the origin of Casimir energy~\cite{casimirobs,casimirobs0} is not the 
vacuum but the energy which is needed to break the symmetry. Fig. \ref{fig:casimir} shows a 
schematic description of these operations.
\begin{wrapfigure}{r}{0.3\textwidth}
\begin{center}
\includegraphics[width=0.3\textwidth]{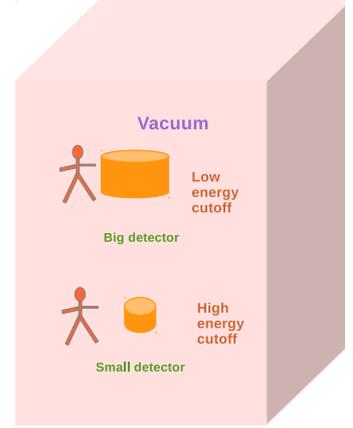} % width is the fraction of text width
\end{center}
\vspace{-1cm}
\caption{\sf \small Schematic description of high energy cutoff on energy. \label{fig:enercutoff}}
%\vspace{-0.5cm}
\end{wrapfigure}
 
We can also interpret the energy remnant from a purely quantum mechanical point of view. A quantum 
field can be decomposed to $\phi (\vec{x},t) = \sum_{\{\alpha\}} u_{\{\alpha\}} (\vec{x},t) a_{\{\alpha\}} + 
u^*_{\{\alpha\}} (\vec{x},t) a_{\{\alpha\}}^\dagger = U (\vec{x},t) + U^\dagger (\vec{x},t),~U (\vec{x},t) 
\equiv \sum_{\{\alpha\}} u_{\{\alpha\}} (\vec{x},t) a_{\{\alpha\}}$. Then 
$\hat{T}^{00} (\vec{x},t)\propto (\dot{U}U^\dagger + \dot{U}^\dagger~U)(\vec{x},t)$. Operators $U$ 
and $U^\dagger$ present annihilation and creation of a particle with any momentum at a spacetime 
point $(\vec{x},t)$. Because the position of the created particle is exactly known, no information 
about its momentum can be obtained, and any value, including infinity, is allowed. These arguments 
are in spirit the same as those given by Eppley \& Hannah~\cite{semiqmgrincon} and 
in~\ref{p9} to prove the inconsistency of a classical gravity and a quantic matter. 
Regularization of the integral in (\ref{t00minkovski}) by imposing a maximum energy cutoff is 
equivalent to considering an uncertainty on the position. It presents the highest energy scale or 
equivalently smallest distances in which the observer can verify the presence of a vacuum and 
provides an upper limit on the vacuum energy density. See Fig. \ref {fig:enercutoff} for a schematic 
description.

\subsection{Contribution of virtual particles and Standard Model condensates in dark energy}
When interactions are considered, the vacuum of quantum field theory is very far from being an 
empty space because quantum fluctuations can condensate~\cite{vaccond}. It is a subject of debate 
whether such cases should be called a {\it vacuum} or not. In any case, we cannot separate the spacetime 
from its quantum field content. For this reason, it is suggested that these condensates 
contribute to vacuum energy and thereby to dark energy~\cite{weinbergde,devacuumrev,devacuumrev0}. 
Even in absence of condensates, loop corrections and running mass and couplings are suggested as 
evidence for coupling of graviton with virtual particles~\cite{devacuumrev,devacuumrev0}, and 
thereby gravitational interaction of vacuum. In this case the density of dark energy should be much 
larger than what is observed. Thus, according to this argument the presence of a small dark energy 
challenges the validity of quantum field theory. Nonetheless, there are several observational 
facts against this criticism which are discussed in detail in~\cite{p2} and can be summarized 
as the following:

After renormalization, the contribution of virtual particles is included in the mass and couplings of 
elementary particles and do not influence large scales. Moreover, renormalization is based on the 
removal of infinities from integrals similar to (\ref{t00minkovski}). The fact that after this 
apparently ad hoc operation we obtain relations that are confirmed by experiments, proves that these 
calculations are after all meaningful, see~\cite{p2} for more evidence. Another indication 
against a gravitational interaction between free particles and vacuum is the stringent constraints on 
the energy dependence of dispersion relation of high energy particles during their propagation over
cosmological distances~\cite{grb090510qgr}. This and other observations put strong constraints on 
the influence of quantum gravity corrections at scales much larger than Planck scale. They also 
constrain the recent suggestion of graviton condensation as the origin of dark 
energy~\cite{quingrcond}, because quantum state of such a condensate would induce fluctuations in 
the propagation of photons proportional to their energy which are not observed. By contrast, there  
is no constraint on the condensate of a field without interaction with visible particles. 

As for the effect of the Standard Model condensates, the most important of them is the newly discovered 
Higgs~\cite{higgsdisco} with a nonzero vacuum expectation value (vev) of $\sim 246$ GeV. 
It is believed to generate mass for the Standard Model particles and triggers the breaking of 
$SU(2) \times U(1)$ symmetry at an scale $\gtrsim 1 TeV$. The conditions for the formation of 
a Bose-Einstein Condensate (BEC) in a quantum fluid is studied in~\cite{beccond}. They demand a 
uniform space distribution for the field. In both classical fluid and quantum field theory the 
amplitude of anisotropies of a condensate decreases very rapidly for large modes. This is analogue to 
the infinite volume condition for symmetry breaking in statistical physics~\cite{beccond}. Nonetheless, 
in presence of additional driver, such as an interaction~\cite{becextforce,p8}, wave 
functions of particles and their condensate are confined to short distances. Therefore, Higgs 
condensate which is coupled to other particles is confined at short distances. In addition, the 
confinement of quarks by QCD helps to confine the Higgs condensate to small scales and it only manifests 
itself through the mass of particles. Similar arguments are used to show that the observed 
pion condensate responsible for the breaking of chiral symmetry is also confined to 
nucleons~\cite{qcdchiralvacu}. In Sec.\ref{sec:deqft-cond} we will show that the survival of the 
quintessence field condensate at cosmological scales is a consequence of its very small mass and very 
weak coupling that leads to the formation of a coherent state which is close to uniform at cosmological 
distances and survives the expansion of the Universe~\cite{p8}.

\subsection{Vacuum as a coherent state}
According to the definition of vacuum in equation (\ref{vacdef}), in the reference frame for which it 
is defined it does not have any particle. Therefore, we expect no effect on a particle 
that passes through vacuum. This property can be used as a test for the presence of a vacuum. However, 
quantum corrections induce an energy dependent effective mass. Therefore, it is the sensitivity 
of an observer to energy variation that determines how well the vacuum can be detected. An observer 
with a high resolution detector never sees any empty space. This means that vacuum is an abstract 
concept. Another issue to consider is the fact that the definition (\ref{vacdef}) 
is not frame independent. However, nonlocality of quantum mechanics and modification of vacuum with 
symmetry breaking raise the necessity for a frame independent definition for the true vacuum of 
quantum field theory. In this section we propose such a definition. 

In~\cite{p8} we have defined a generalized coherent state $|\Psi_{GC}\rangle$ for a scalar 
field based on an original suggestion by~\cite{cohereglauber,condwave}:
\bea
&& |\Psi_{GC}\rangle \equiv \sum_k A_k e^{C_k a_k^{\dagger}} |0\rangle = \sum_k A_k 
\sum_{i=0}^{N \rightarrow \infty} \frac {C_k^i}{i!}(a_k^{\dagger})^i |0\rangle \label{condwaveg} \\ 
&& a_k |\Psi_{GC}\rangle = C_k |\Psi_{GC}\rangle \quad \quad 
\langle \Psi_{GC}| N_k |\Psi_{GC}\rangle = |A_k C_k|^2 \label{condwavegann}
\eea
For $\{C_k \rightarrow 0~\forall~k\}$ this state is neutralized by all annihilation operators and 
the expectation value of particle number approaches zero for all modes. Therefore, this state 
satisfies the condition (\ref{vacdef}) for a vacuum state. Coefficients $A_k$ are relative 
amplitude of modes and can be nonzero even for a vacuum state. In addition, one can extend this 
definition by considering products of $|\Psi_{GC}\rangle$ states. Such a state includes products of 
states in which particles do not have the same momentum, thus it consists of all combinations of 
states with any number of particles and momenta: 
\be
|\Psi_{G}\rangle \equiv \sum_{k_1,k_2,\cdots} \biggl (\prod_{k_i} A_{k_i} \biggr) e^{\sum_i C_{k_i} 
a_{k_i}^{\dagger}} |0\rangle \label{condwave}
\ee
A vacuum state is defined as $C_{k_i} \rightarrow 0~\forall~k_i$ which is asymptotic limit of 
non-vacuum states. Under a Bogoliubov transformation this state is projected to itself:
\be
a_{k_i} = \sum_j \sum_{k_j} {\mathcal A}_{k_j k_i} a'_{k_j} + \sum_j \sum_{k_j} 
{\mathcal B}_{k_j k_i} {a'}^\dagger_{k_j} \quad \quad a^\dagger_{k_i} = \sum_j \sum_{k_j} 
{\mathcal A}^*_{k_j k_i} {a'}^\dagger_{k_j} + \sum_j \sum_{k_j} {\mathcal B}^*_{k_j k_i} a'_{k_j}
\label {bogoltrans}
\ee
Replacing $a_{k_i}^{\dagger}$ in (\ref{condwave}) with the corresponding expression in 
(\ref{bogoltrans}) leads to an expression for $|\Psi_{G}\rangle$ similar to (\ref{condwave}) but 
with respect to the new operator ${a'}^\dagger_k$ and $C'_{k_j} = \sum_i \sum_{k_i} 
{\mathcal A}^*_{k_j k_i} C_{k_i}$. For $C_{k_i} \rightarrow 0~\forall~k_i$ and finite 
${\mathcal A}^*_{k_j k_i}$, $C'_{k_i} \rightarrow 0~\forall~k_i$. Note that here we assume that the 
Bogoliubov transformation changes $|0\rangle$ to a similar state which is neutralized by 
$a'_k~\forall~k$. Therefore, in contrast to the null state of the Fock space, $|\Psi_{G}\rangle$ 
is frame-independent. 

It is easy to verify that this new definition of vacuum does not solve the problem of singularity 
of $\hat{T}^{00}$ expectation value. Nonetheless, it gives a better insight into the nature of the 
problem. Notably, one can use the number operator $\sum_k \hat{N}_k$ to determine the energy density 
of vacuum because in contrast to $|0\rangle$, the new vacuum $|\Psi_{G}\rangle$ is frame-independent 
and is neutralized by the number operator $\hat {N}_k |\Psi_{G}\rangle = 0 ~\forall~k$. This 
alternative to $\hat{T}^{00}$ for measuring the vacuum energy density has been discussed 
in~\cite{qftcurve}, but has been considered to be a poor replacement because the vacuum state 
defined in (\ref{vacdef}) is not frame independent. Note that we explicitly distinguish between 
a system in which all particles are in the ground state and a system in which the expectation 
value of particle number in any state, including the ground state, is zero. We call the first 
system a condensate and the second one according to (\ref{condwavegann}) is a vacuum. Therefore, 
according to this definition string vacua of modulies after compactification are condensates. 
See~\cite{p2} for more examples and details.

Although coefficients $A_k$ which must be calculated from the full Lagrangian depend on the initial 
or boundary conditions, the state $|\Psi_{G}\rangle$ contains all combinations of particles and is 
always projected to itself when the reference frame is changed. In this sense it is a unique maximally 
coherent state. Like any superposition state its observation - which needs an interaction - leads to 
a collapse to one of its member states. An external observer interprets this as observation of virtual 
particles - because they come from a presumed vacuum - and their 
effect manifests itself as scale dependence of mass and couplings of the field. Because any single 
state in the vacuum superposition has a vanishing amplitude, one can always consider that the vacuum 
stays unchanged even when one or any finite number of its members interact and decohere. Thus, like 
usual definition of vacuum, interactions modify properties of the external (untangled, on-shell) 
particles at scales relevant for their interaction, but they do not change $|\Psi_{G}\rangle$ globally. 
\begin{wrapfigure}{r}{0.3\textwidth}
\vspace{-1.cm}
\begin{center}
\includegraphics[width=0.3\textwidth]{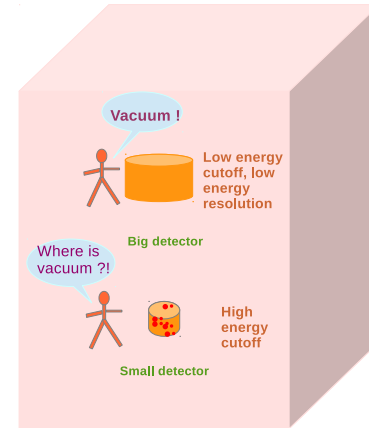} % width is the fraction of text width
\end{center}
\vspace{-1.cm}
\caption{\sf \small A schematic description of production of virtual particles by partial 
decoherence of vacuum. The points (red) present virtual particles which are decohered and their 
indirect effects can be measured by detectors as scale-dependence of mass and couplings. 
\label{fig:novacuum}}
\vspace{-1.cm}
\end{wrapfigure}
The vacuum $|\Psi_{G}\rangle$ includes all states at any scale, but in every experiment only 
a range of them are available to observers. They are limited from IR side by the size of 
the apparatus or observational limits such as a horizon, and from UV side by the available energy 
to the observer. The presence of a particle at a given scale i.e. discrimination between vacuum 
and non-vacuum at that scale depends on the uncertainties in distance/energy measurements. At large 
distance scales the limited sensitivity of detectors cannot detect interaction with very low energy 
virtual particles, thus no violation of energy-momentum conservation occurs. This could not be true 
if the vacuum had a large energy-momentum density which could be exchanged with on-shell particles.

Does the coherent vacuum state $|\Psi_{G}\rangle$ gravitate ? A detailed answer to this question 
needs a quantum description for gravity. Nonetheless, by definition states that make up a coherent 
state are not observable except when they are decohered/collapsed. And when this happens, they will 
no longer appear as vacuum, see Fig. \ref{fig:novacuum} for a schematic illustration. Therefore, 
they cannot influence observations in any way, including gravitationally. In a semi-classical view, 
one expects that the expectation value of the number of particles with a given energy and momentum 
determines the strength of the gravitational force. Equation (\ref{condwavegann}) shows that this 
number for any mode $k$ is null when $\{C_k \rightarrow 0~\forall~k\}$. Thus, this state does not 
feel the gravity. This is another evidence of the unphysical nature of the singularity of the 
expectation value of energy-momentum tensor when its classical definition is used in quantum field 
theory without any regularization. 

\subsubsection{Vacuum or not Vacuum}
A question which arises here is: Why does the existence of a frame-independent vacuum rule out 
vacuum energy as the origin of accelerating expansion ? The answer to this question depends on 
what we mean by vacuum. If by vacuum we mean the minimum of the effective classical - condensate - 
component of quantum fields, then dark energy may be considered as the vacuum energy. However, as 
we discussed in this section and will demonstrated in details in Sec. \ref {sec:deqft-cond}, 
the condensate is very far from being the particle-less state that conventional word {\it vacuum} 
means and the quantum state used in (\ref {vacdef}) designates. Therefore, minimums of the 
condensate potential should not be called {\it vacuum}. Moreover, we will show that we can begin 
from an initial moment where the amplitude of a condensate is zero, i.e. the state of the Universe 
is $|\Psi_{G}\rangle$ with $\{C_k \rightarrow 0~\forall~k\}$ and study its formation and 
evolution. These processes are also very far from the static concept of vacuum. In laboratory 
condensed matter the lowest energy state is usually called {\it vacuum} as synonymous to the 
background state above which - usually in energetic sense - fluctuations are studied. In the 
context of early Universe physics, the {\it background} state itself is as important and unknown 
as its fluctuations. Thus, employment of ambiguous terms such as {\it vacuum} only adds to 
confusion.

\subsection{Outline}
In this section, various arguments were put forward to advocate a null energy density for vacuum in 
the context of quantum field theories. They rule out the energy density of vacuum as the origin of 
dark energy. The vacuum state was shown to be an abstract concept that only approximately and 
asymptotically makes sense. We proposed a new frame-independent definition for vacuum as a coherent 
state with an amplitude approaching zero. Apart from helping to understand issues regarding the 
origin of dark energy, this definition may be useful for nonlocal description of quantum gravity 
and systems including condensates. In absence of a vacuum energy in the sense we defined here, 
the best candidates for dark energy are modification of the Einstein gravity and condensation of 
one or multiple quantum fields with quintessential behaviour. 

\section{Quintessence and interaction in the dark sector} \subsectionmark{Quintessence and interaction $\cdots$} \label{sec:deqft-int} % ~5 pages
\subsection{Introduction}
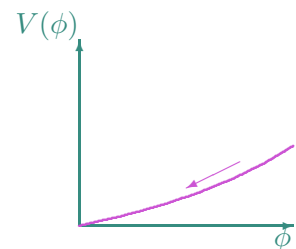
\begin{wrapfigure}{r}{0.3\textwidth}
\vspace{1cm}
\begin{center}
{\color{darkturquoise}
\begin {picture}(75,35)(-10,0)
\linethickness{0.8pt}
\put(0,0){\vector(1,0){80}}
\put(0,0){\vector(0,1){70}}
{\color{mediumorchid}
\qbezier(0,0)(50,10)(80,30)
\put(60,24){\vector(-2,-1){20}}}
\put(76,-10){\makebox(0,0)[b]{$\phi$}}
\put(-12,70){\makebox(0,0)[b]{$V(\phi)$}}
\end {picture}
}
\caption{\sf \small Rolling down of a quintessence field beginning in the early Universe and lasting up 
to now. \label{fig:quinpot}}
\end{center}
\end{wrapfigure}
Even before the observational confirmation of an entity in the Universe behaving very similar 
to a cosmological constant, cosmologists have tried to find models with such behaviour at recent 
epochs in the history of the Universe~\cite{quin}. These models are generically called 
{\it Quintessence} and majority of them are based on one or more scalar fields, although models 
based on vector fields have been also studied~\cite{quinvector}. Similar to slow-roll models of 
inflation, the scalar field asymptotically approaches to the minimum of the potential at zero. In a 
variant of quintessence models called {\it k-essence}~\cite{kessence} the evolution of the scalar 
field is governed by the kinetic energy which has a non-standard form generally written as 
$K = f (\phi, \partial_\mu \phi)$. These models are usually inspired from string and other quantum 
gravity models. In some quintessence models it is considered that the same field that has generated 
inflation in the early Universe plays also the role of quintessence at present~\cite{infquinunite}, 
see e.g.~\cite{infquinunite0} for comparison with recent data.

It is not a trivial task to make models with what is called {\it a tracking solution}, which without 
fine-tuning of parameters and initial conditions for a duration more than half of the age of the 
Universe approaches zero without reaching to this limit point. It is shown~\cite{quin,trackingcond} 
that the necessary condition for the presence of such solutions is: 
\be
V"V/V'^2 > 1 \label{trackcond}
\ee
It is easy to verify that for analytically simple models with polynomial or exponential potentials, 
the condition (\ref {trackcond}) is satisfied if their order or exponent is negative. Such potentials 
and k-essence models are not renormalizable except when the model is linearized and only small 
fluctuations are quantized~\cite{quinquantum}. For this reason they must be considered as effective 
models. On the other hand, it is expected that a quintessence field has very weak interactions, thus 
its effective potential must be close to its bare potential and perturbative. This conclusion is not 
consistent with a nonperturbative potential.

Apart from non-renormalizability several other issues about quintessence models with only 
self-interaction can be remarked. As it is described in section \ref{sec:deqft-intro}, in these 
models the equation of state of dark energy:
\be
w = \frac {P}{\rho} = \frac {\frac{1}{2}\dot{\phi}^2 - V (\phi)}{\frac{1}{2}\dot{\phi}^2 + 
V (\phi)} \geqslant -1 \label{wdefquin}
\ee
does not allow a phantom-like behaviour for positive value of the potential, and may be 
inconsistent with data~\cite{quinnewobs}. Phantom models correspond to a Wick rotation of time 
coordinate in ordinary quintessence models, and are their Euclidean analogues~\cite{quinphantom}. 
Under this operation $\rho \rightarrow -P$, $P \rightarrow - \rho$, 
and $w \rightarrow 1/w < -1$. Although the Wick rotation technique is used in many circumstances in 
physics for simplifying calculations, usually an inverse rotation is performed at the end to bring 
back calculations to Lorentzian metric. In phantom models the Wick rotation is performed only in 
quintessence sector and it is not rotated back. Thus, it is considered that the field is in this 
state only for a limited time. In any case such a model can be only an effective toy model. 

In addition to $w \geqslant -1$ issue, simple/pure quintessence models and a cosmological constant 
do not explain why the density of dark energy is fine-tuned such that galaxies could be formed before 
it becomes dominant. This problem is called {\it dark energy coincidence}~\cite{weinbergde}. 
We should remind that if the origin of dark energy is related to physics at Planck scale - as in the 
context of string theory - or even at lower scales such as inflation era or reheating, its density 
was tens of orders of magnitude smaller than matter density at early epochs. This needs an 
extreme fine-tuning unless there is an inherent relation between dark energy and other constituents 
of the Universe, for instance through an interaction between quintessence field and dark 
matter~\cite{quinint,p29,p25,p23}. Another advantage of this class of models is that they provide 
a natural explanation for $w_{obs} < -1$ if the interaction of dark energy with matter is ignored 
in analysing the data~\cite{p29,quinintw}. 

We should remind that quintessence models studied in pioneering works~\cite{quinint} are actually 
modified gravity related to Brans-Dick extension of Einstein gravity. A dilaton scalar field is 
introduced through the transformation $g_{\mu\nu} \rightarrow e^{C\phi}g_{\mu\nu}$ in matter Lagrangian 
where $\phi$ is the quintessence/dilaton field, $g_{\mu\nu}$ is the metric, and $C$ a coupling constant. 
In~\cite{quinint0} the couplings of dilaton to dark and baryonic matter are different. They reflect 
different conformal properties of these constituents and create a fifth force effect which may induce 
a segregation between these two types of matter. Another class of interacting quintessence models which 
have been extensively studied are scaling 
models~\cite{quinscaling0,quinscaling1,quinscaling2,quinscaling3}. They assume a 
constant ratio of dark matter and dark energy. However, their prediction for the equation of 
state is $w \gtrsim -0.7$ which is inconsistent with data. Other type of scaling models based on 
non-standard kinetic term inspired or associated to string theory and/or brane models are also 
proposed~\cite{quindisformal}. In this class of models, which can be classified as k-essence, the 
non-standard expression of the kinetic term is due to induced metric of a 4D brane boundary by a 
dilaton field living in the higher dimensional bulk. Thus, the effective quintessence field on the 
visible brane has a geometric/gravitational origin. In our knowledge the model presented 
in~\cite{p25,p20} is the first properly speaking interacting quintessence model with a particle 
physics interpretation rather than geometry.

Many extensions of the 
Einstein gravity and classical limit of quantum gravity models include a scalar field. The best 
example is $F(R)$~\cite{modgrfr} and conformal gravity. In these models the scalar field usually 
has a non-minimal interaction with matter, and therefore apparently they are similar to 
interacting quintessence. Nonetheless, we can in principle distinguish them from their interaction, 
although it is a very challenging task, see Sec. \ref{sec:deqft-param}. If like gravity, the scalar 
field(s) has(have) similar coupling to all type of matter, we call it 
modified gravity, otherwise an interacting quintessence. However, in an observational point of view 
it would be very difficult to test this criterion, because the scalar field is expected to have a 
very weak interaction with matter. Moreover, about $80 \%$ of matter in the Universe is dark and 
cannot be observed directly. Consequently, the measurement of difference between coupling of dark 
energy to various component is very difficult if not impossible. For this reason other discriminating 
criteria should be used. This point will be discussed in detail in section \ref{sec:deqft-param}.

\subsection{Dark energy with $w_{obs} \lesssim -1$ in presence of interaction in the dark sector} 
\subsectionmark{Dark energy with $w_{obs} \lesssim -1~~\cdots$}
Strangely enough my involvement in dark energy research began during the study of Ultra High Energy 
Cosmic Rays (UHECRs)! This study is summarized in Sec. \ref{sec:he-uhecr}. Here we just mention that 
to verify the consistency of a decaying super-heavy dark matter as the origin of UHECRs in the 
context of top-down models, in~\cite{p29} the evolution of Hubble function $H(z)$ in these models is 
compared with the available set of data from supernovae type Ia~\cite{snmeasur,snpro,snobs} for 
various lifetime of dark matter. In a flat universe with a cosmological constant $H(z)$ is:
\be
H^2 (z) = \frac {8\pi G}{3} T^{00} (z) + \frac {\Lambda}{3} \label {hz}
\ee
where $T^{\mu\nu}$ is the energy-momentum tensor. For stable matter and radiation $T^{00}(z) = 
\rho_c \Omega_m (1+z)^3 + \Omega_h (1+z)^4$ where $\rho_c \equiv 3H_0^2/8\pi G$ is the present critical 
density, $\Omega_m$ and $\Omega_h$ are fractional density of cold and hot matter, respectively, and 
$z$ is the redshift. When dark matter decays or interacts with other components~\cite {quinintw}, 
there is no exact expression for $T^{00}$ because it depends on the decay modes of the meta-stable 
dark matter, its elastic and non-elastic couplings, and the fate of remnants i.e. whether they 
are and stay relativistic or lose their energy and become non-relativistic. For this reason 
numerical simulations  of the decay of a super-heavy dark matter which includes the propagation 
and dissipation of its remnants is used to determine $T^{00}$.

Fig. \ref{fig:bestres} shows the best fit of data with simulations. The data used in~\cite{p29} 
is the published data set B of the Supernova Cosmology Project~\cite{snmeasur,snproj,snproj0} for 
high redshift and Calan-Tololo sample~\cite{tololo} for low redshift supernovae. It should be 
reminded that $w_m$ the equation of state of matter for a decaying/interacting dark matter is not 
null, i.e. it is not an ideal Cold Dark Matter (CDM) with $w_m = 0$. This is an important point 
because it is this difference that leads to equation state of dark energy $w_{de} \lesssim -1$ if the 
decay or interaction of dark matter is not taken into account in the model fitted to the data. 
We call the value of $\Omega_\Lambda$ and $w_{de}$ derived with the assumption of a stable dark matter 
$\Omega_{\Lambda}^{eq}$ and $w_{de}^{eq}$. These values should be compared with what is in the literature 
because their null hypothesis is usually a stable CDM. 
\begin{figure}
\begin{center}
\includegraphics[width=10cm]{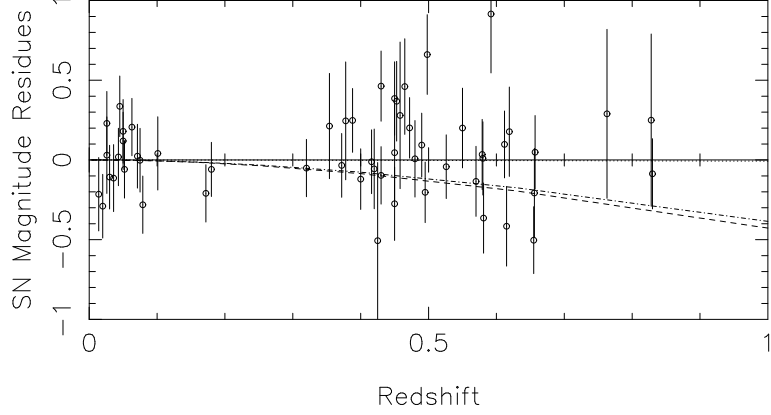}
\caption{\sf \small Best fit residues of apparent magnitude of supernovae for 
$\Omega_{\Lambda}^{eq} = 0.7$, $\tau = 5 \tau_0$, where $\tau_0$ is the present age of the Universe and 
$\tau$ the lifetime of dark matter in simulations. The value of fractional cosmological constant 
density used in this simulation is $\Omega_{\Lambda} = 0.73$ (full line). Other curves are for stable 
dark matter with $\Omega_{\Lambda}^{eq} = \Omega_{\Lambda} = 0.7$ (doted); decaying dark matter with 
$\tau = 5 \tau_0$ and $\Omega_\Lambda = 0$ (dash line); stable DM and $\Omega_\Lambda = 0$ (dash-dot). 
\label{fig:bestres}}
\end{center}
\end{figure}
Additionally, for each model of decaying dark matter and a cosmological constant as dark energy we 
determined parameters of an equivalent quintessence model with stable dark matter and density 
$\rho_q = \Omega_q (1+z)^{3(w_q + 1}$. Fig. \ref {fig:quindata} shows the variation of $\chi^2$ of 
the fit with $w_q$. Table \ref{tab:quineq} shows the value of parameters for the equivalent 
quintessence models. Regarding the value of $\chi^2$ of these models, they are all consistent with 
data. However, clearly models with $w_q < -1$ fit the data somehow better, except the model with 
$\Omega_\Lambda = 0.8$ which has a poorer fit than others. {\bf This proves that the wrong priory 
of a stable or non-interacting matter can lead to $w_q < -1$ when dark matter decays or interacts 
and dark energy is a cosmological constant.} This work is one of the first work in which it was 
shown that the observed $w < -1$ for dark energy can be due to the application of a wrong model. 
Results shown in Fig. \ref{fig:quindata} and Table \ref{tab:quineq} are consistent with latest 
measurements of the equation state of dark energy from supernova data~\cite{snlatest}, LSS, 
and CMB anisotropies~\cite{quinnewobs,}.  
\begin{figure}
\begin{center}
\includegraphics[width=12cm]{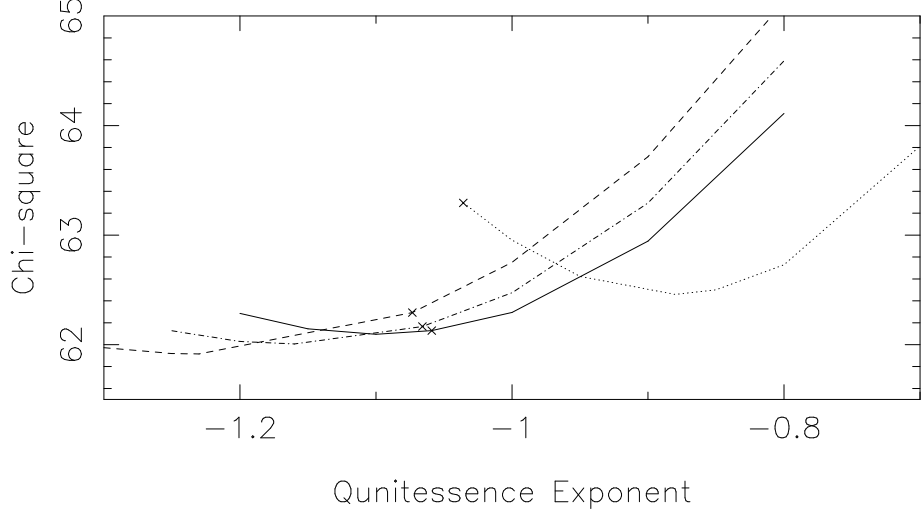}
\caption{\sf \small $\chi^2$-fit of supernovae data with quintessence models as a function of $w_q$ 
for $\Omega_q = 0.67$ (dashed), $\Omega_q = 0.69$ (dash-dot), $\Omega_q = 0.71$ (solid) and 
$\Omega_q = 0.8$ (dotted). $\chi^2$'s of equivalent quintessence models to a decaying dark matter 
with $\tau = 5 \tau_0$ and closest $\Omega_\Lambda$ to $\Omega_q$ (see table \ref{tab:quineq}) are 
also shown. $\tau$ and $\tau_0$ are respectively the lifetime of dark matter in simulations and 
the present age of the Universe. Except $\Omega_q = 0.8$ model others are all good fits to data. 
For $\Omega_\Lambda = \Omega_q = 0.8$, a stable DM fits the data better, but the fit is poorer than 
former models.\label{fig:quindata}}
\end{center}
\end{figure}
\begin{table}
\vspace{0.2cm}
\begin{center}
\footnotesize
\begin{tabular}{|c|c|c|c|c|c|c|c|c|c|}
\hline
 &
\multicolumn {3}{c|}{Stable DM} & 
\multicolumn {3}{c|}{$\tau = 50 \tau_0$} & 
\multicolumn {3}{c|}{$\tau = 5 \tau_0$} \\
\hline
 &
\raisebox{0pt}[13pt][7pt]{$\Omega_{\Lambda} = 0.68$} &
\raisebox{0pt}[13pt][7pt]{$\Omega_{\Lambda} = 0.7$} &
\raisebox{0pt}[13pt][7pt]{$\Omega_{\Lambda} = 0.72$} &
\raisebox{0pt}[13pt][7pt]{$\Omega_{\Lambda} = 0.68$} &
\raisebox{0pt}[13pt][7pt]{$\Omega_{\Lambda} = 0.7$} &
\raisebox{0pt}[13pt][7pt]{$\Omega_{\Lambda} = 0.72$} &
\raisebox{0pt}[13pt][7pt]{$\Omega_{\Lambda} = 0.68$} &
\raisebox{0pt}[13pt][7pt]{$\Omega_{\Lambda} = 0.7$} &
\raisebox{0pt}[13pt][7pt]{$\Omega_{\Lambda} = 0.72$}\\
\hline
\raisebox{0pt}[12pt][6pt]{$H_0$}  
 & \raisebox{0pt}[12pt][6pt]{$69.953$}
 & \raisebox{0pt}[12pt][6pt]{$69.951$} & \raisebox{0pt}[12pt][6pt]{$69.949$}
 & \raisebox{0pt}[12pt][6pt]{$69.779$} & \raisebox{0pt}[12pt][6pt]{$69.789$}
 & \raisebox{0pt}[12pt][6pt]{$69.801$} & \raisebox{0pt}[12pt][6pt]{$68.301$}
 & \raisebox{0pt}[12pt][6pt]{$68.415$} & \raisebox{0pt}[12pt][6pt]{$68.550$}\\
\hline
\raisebox{0pt}[12pt][6pt]{$\Omega_{\Lambda}^{eq}$}
 & \raisebox{0pt}[12pt][6pt]{$0.681$}
 & \raisebox{0pt}[12pt][6pt]{$0.701$} & \raisebox{0pt}[12pt][6pt]{$0.721$}
 & \raisebox{0pt}[12pt][6pt]{$0.684$} & \raisebox{0pt}[12pt][6pt]{$0.704$}
 & \raisebox{0pt}[12pt][6pt]{$0.724$} & \raisebox{0pt}[12pt][6pt]{$0.714$}
 & \raisebox{0pt}[12pt][6pt]{$0.733$} & \raisebox{0pt}[12pt][6pt]{$0.751$}\\
\hline
\raisebox{0pt}[12pt][6pt]{$\Omega_q$}
 & \raisebox{0pt}[12pt][6pt]{-}
 & \raisebox{0pt}[12pt][6pt]{-} & \raisebox{0pt}[12pt][6pt]{-}
 & \raisebox{0pt}[12pt][6pt]{$0.679$} & \raisebox{0pt}[12pt][6pt]{$0.700$}
 & \raisebox{0pt}[12pt][6pt]{$0.720$} & \raisebox{0pt}[12pt][6pt]{$0.667$}
 & \raisebox{0pt}[12pt][6pt]{$0.689$} & \raisebox{0pt}[12pt][6pt]{$0.711$}\\
\hline
\raisebox{0pt}[12pt][6pt]{$w_q$}
 & \raisebox{0pt}[12pt][6pt]{-}
 & \raisebox{0pt}[12pt][6pt]{-} & \raisebox{0pt}[12pt][6pt]{-}
 & \raisebox{0pt}[12pt][6pt]{$-1.0066$} & \raisebox{0pt}[12pt][6pt]{$-1.0060$}
 & \raisebox{0pt}[12pt][6pt]{$-1.0055$} & \raisebox{0pt}[12pt][6pt]{$-1.0732$}
 & \raisebox{0pt}[12pt][6pt]{$-1.0658$} & \raisebox{0pt}[12pt][6pt]{$-1.0590$}\\
\hline
\raisebox{0pt}[12pt][6pt]{$\chi^2$}
 & \raisebox{0pt}[12pt][6pt]{$62.36$}
 & \raisebox{0pt}[12pt][6pt]{$62.23$} & \raisebox{0pt}[12pt][6pt]{$62.21$}
 & \raisebox{0pt}[12pt][6pt]{$62.34$} & \raisebox{0pt}[12pt][6pt]{$62.22$}
 & \raisebox{0pt}[12pt][6pt]{$62.21$} & \raisebox{0pt}[12pt][6pt]{$62.22$}
 & \raisebox{0pt}[12pt][6pt]{$62.15$} & \raisebox{0pt}[12pt][6pt]{$62.20$}\\
\hline
\end{tabular}
\end{center}
\caption{\sf \small Cosmological parameters from simulations of a decaying DM and parameters of 
the equivalent quintessence models. $H_0$ is in $km$ $Mpc^{-1}\sec^{-1}$.\label{tab:quineq}}
\end{table}

It is well known that simulations and fitting include many approximations and uncertainties. To prove 
that the conclusion about the effect of neglecting decay/interaction of dark matter on the measured 
value of $w$ is not an artifact, in~\cite{p29} an approximate analytical demonstration was also 
performed. Due to the importance of this demonstration for our understanding of 
observations and models it is summarized here: 

\subsubsection{Analytical demonstration} With a good precision the total density of a decaying dark 
matter model can be written as the following:
\be
\frac {\rho (z)}{{\rho}_c} \approx {\Omega}_m (1 + z)^3 \exp 
(\frac {{\tau}_0 - t}
{\tau}) + {\Omega}_{hot} (1 + z)^4 + {\Omega}_m (1 + z)^4 \biggl (1 - 
\exp (\frac {{\tau}_0 - t}{\tau}) \biggr ) + {\Omega}_{\Lambda}. \label {totdens}
\ee
where $\rho_c$ is the critical density of the Universe at redshift zero and ${\Omega}_{hot}$ 
is the fractional density of relativistic components. It is assumed that decay remnants are relativistic 
particles and their dissipation is neglected. In a flat cosmology 
${\Omega}_m + {\Omega}_{hot} + {\Omega}_{\Lambda} = 1$ and ${\rho}_c$ is the present critical density. 
If dark matter is stable and we neglect the contribution of hot dark matter, the expansion factor 
$a(t)$ is:
\be
\frac {a (t)}{a ({\tau}_0)} = \biggl [ \frac {(B \exp (\alpha (t - 
{\tau}_0)) - 1)^2}{4AB \exp (\alpha (t - {\tau}_0))}\biggr ]^{\frac {1}{3}} 
\equiv \frac {1}{1 + z}. \label {at}
\ee
\bea
A & \equiv & \frac {{\Omega}_{\Lambda}}{1 - {\Omega}_{\Lambda}}, \\
B & \equiv & \frac {1 + \sqrt {{\Omega}_{\Lambda}}}{1 - 
\sqrt {{\Omega}_{\Lambda}}}, \\
\alpha & \equiv & 3 H_0 \sqrt {{\Omega}_{\Lambda}}.
\eea
Using (\ref {at}) as an approximation for $\frac {a (t)}{a ({\tau}_0)}$ when dark matter slowly decays, 
(\ref {totdens}) takes the following form:
\bea
\frac {\rho (z)}{{\rho}_c} & \approx & {\Omega}_m (1 + z)^3 C^{-\frac {1}{\alpha \tau}} + 
{\Omega}_{hot} (1 + z)^4 + {\Omega}_m (1 + z)^4 (1 - C^{-\frac {1}{\alpha \tau}}) + 
{\Omega}_{\Lambda}. \label {totdens1}\\
C & \equiv & \frac {1}{B} \biggl (1 + \frac {4A}{(1 + z)^3} - 
\sqrt {(1 + \frac {4A}{(1 + z)^3})^2 - 1} \biggr).
\eea
For a slowly decaying dark matter, $\alpha \tau \gg 1$ and (\ref {totdens1}) becomes:
\bea
\frac {\rho (z)}{{\rho}_c} & \approx & {\Omega}_m (1 + z)^3 + {\Omega}_{hot} (1 + z)^4 + 
{\Omega}_q (1 + z)^{3 {\gamma}_q}, \label {totdens2} \\
{\Omega}_q (1 + z)^{3 {\gamma}_q} & \equiv & {\Omega}_{\Lambda} (1 + 
\frac {{\Omega}_m}{\alpha \tau {\Omega}_{\Lambda}} z (1 + z)^3 \ln C). 
\label {qeqdef}
\eea
Equation (\ref {qeqdef}) is the definition of equivalent quintessence component. After its 
linearization:
\be
w_q \equiv {\gamma}_q - 1 \approx \frac {{\Omega}_m (1 + 4 A)(1 - \sqrt {2 A})}
{3 \alpha \tau {\Omega}_{\Lambda} B} - 1.
\ee
It is easy to see that in this approximation $w_q < -1$ if ${\Omega}_{\Lambda} > \frac {1}{3}$.

\subsection{Quintessence from decay of a super-heavy dark matter}
Motivated by the arguments given in section \ref{sec:deqft-int} in favour of an interacting 
quintessence model, a model with a slowly disintegrating dark matter is considered which has a very 
weak interaction with a light scalar considered to be the quintessence field~\cite {p25}. 
Furthermore, through the study of this class of models, some of advantages of an interaction in the 
dark sector are shown. 

Consider that just after inflation the Universe consists of a cosmological {\it soup} including 2 
species: a superheavy dark matter (SDM) called $X$ which is decoupled from the rest of the 
soup since very early times, and the ensemble of other species which we do not specify in detail. 
The only constraint imposed on the latter component is that it must consist of light species in 
- comparison with $X$ - including: baryons, neutrinos, photons, and light dark matter. For 
simplicity we assume that $X$ is a scalar field $\phix$ and is meta-stable, i.e. it decays with a 
lifetime much longer than the present age of the Universe. A very small energy fraction of the decay 
remnants is transferred to a light scalar quintessence field $\phiq$ with negligibly weak interaction 
with other fields. This model is motivated by top-down models for the origin of Ultra High Energy 
Cosmic Rays~\cite{defect,hanti1,p64,p63,p59,p27,p23}. 
Despite recent arguments against this type of models~\cite{uhecragncorr,uhecrorigrev}, they are not 
yet completely ruled out. And even if it turns up that UHECRs have astronomical origin, dark matter or 
one of its constituents can be meta-stable, see e.g.~\cite{crypton1,uhecrorigrev}. Remind that smaller 
the fraction of this type of particles in the dark matter, shorter their lifetime and larger their 
coupling to the quintessence field are allowed~\cite{p27}.

The effective Lagrangian of this model is:
\be
{\mathcal L} = \int d^4 x \sqrt{-g} \biggl [\frac {1}{2} g^{\mu\nu} 
{\partial}_{\mu} \phix {\partial}_{\nu} \phix + \frac {1}{2} g^{\mu\nu} 
{\partial}_{\mu} \phiq {\partial}_{\nu} \phiq - V (\phix, \phiq, J) 
\biggr ] + {\mathcal L}_J \label {lagrange}
\ee
The field $J$ presents collectively light particles. The term $V (\phix, \phiq, J)$ comprises all 
interactions including self-interaction potentials for $\phix$ and $\phiq$:
\be
V (\phix, \phiq, J) = V_q (\phiq) + V_x (\phix) + g {\phix}^m {\phiq}^n + 
W (\phix, \phiq, J) \label {potv}
\ee
The term $g {\phix}^m {\phiq}^n$ is important because it is responsible for the annihilation of $X$ 
and back reaction of quintessence field. $W (\phix, \phiq, J)$ presents other interactions which 
contribute to decay of $X$ to light fields $J$ and $\phiq$. To determine the evolution of 
these fields we consider $X$ and $J$ as classical particles. The contribution of quintessence field 
$\phiq$ consists of classical relativistic particles with density $\rho'_q$ and a condensate 
component behaving as dark energy with density $\rho_q$. Under these simplifying assumptions, the 
evolution equations of various components of the model are written as the followings:
\bea
\dot{\phiq} [\ddot{\phiq} + 3H \dot{\phiq} + {m_q}^2 \phiq + 
\lambda {\phiq}^3] & = & -2g \dot{\phiq}\phiq 
\biggl (\frac {2 {\rho}_x}{{m_x}^2}\biggr ) + {\Gamma}_q{\rho}_x 
\label {phiqe} \\
\dot {{\rho}_x} + 3H {\rho}_x & = & - ({\Gamma}_q + {\Gamma}_J)
{{\rho}_x} - {\pi}^4 g^2 \biggl (\frac {{{\rho}_x}^2}{{m_x}^3} - 
\frac {{{\rho}_q}'^2}{{m_q}^3}\biggr ) \label {xeq} \\
\dot {{\rho}_J} + 3H ({\rho}_J + P_J) & = & {\Gamma}_J {{\rho}_x} 
\label {jeq} \\
H^2 & \equiv & \biggl (\frac {\dot{a}}{a}\biggr )^2 = \frac {8\pi G}{3} 
({\rho}_x + {\rho}_J + {\rho}_q) \label {heq} \\
{\rho}_q & = & \frac {1}{2} {m_q}^2 {\dot{\phiq}}^2 + 
\frac {1}{2} {m_q}^2 {\phiq}^2 + \frac {\lambda}{4} {\phiq}^4 
\label {phidens}
\eea
The constants ${\Gamma}_q$ and ${\Gamma}_J$ are respectively the decay width of $X$ to $\phiq$ and to 
other species. The effect of decay term $W (\phix, \phiq, J)$ in the Lagrangian appears as the 
total decay rate of $X$ particles $({\Gamma}_q + {\Gamma}_J){\rho}_x$ in energy conservation 
equation. The effect of $g$-coupling is considered separately.

The system of equations (\ref {phiqe})-(\ref {phidens}) is highly non-linear and an analytical 
solution cannot be found easily. There are however two asymptotic regimes which permit an approximate 
analytical treatment. The first solution corresponds to early times just after the production of $X$ 
particles, presumably after 
preheating~\cite {wimpzilla})~\cite{decoherethermal,decohereinf0,decohereinf1}. In 
this epoch $\phiq \sim 0$ and can be neglected. The other regime is when the time variation of 
$\phiq$ becomes very slow and one can neglect $\ddot {\phiq}$. In~\cite{p25} it is shown that 
these regimes can be connected smoothly and the final solution is very close to a constant, i.e. the 
quintessence field imitates a cosmological constant.

Numerical solution of equations (\ref{phiqe}) to (\ref{phidens}) confirms the above approximate 
analytical conclusions. Fig. \ref {fig:quinevol} presents the results of the numerical calculation 
of the evolution of the density of quintessence field in this type of models. We conclude that for a 
large fraction of the parameter space and without fine-tuning, the scalar field varies very slowly 
soon after beginning of its formation, and behaves similar to a cosmological constant.
\begin{figure}[h]
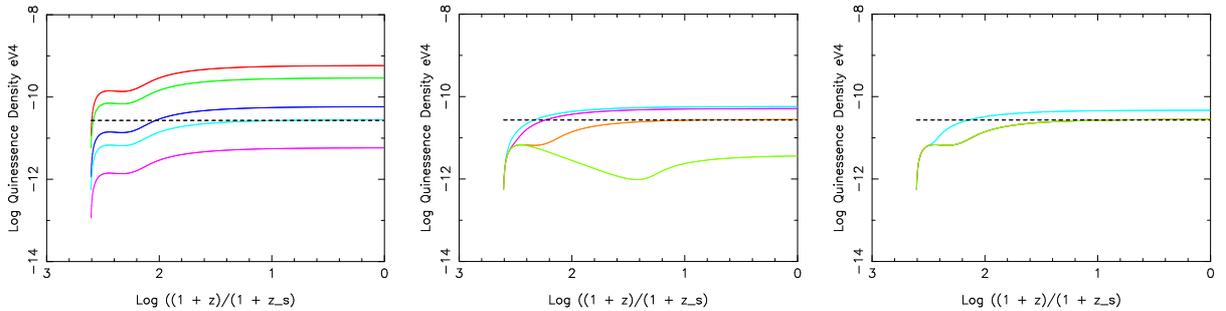

\begin{center}
\begin{tabular}{lll}
\includegraphics[width=4cm,angle=-90]{quindens.eps} &
\includegraphics[width=4cm,angle=-90]{quindenscompmass.eps} &
\includegraphics[width=4cm,angle=-90]{quindenscomplambda.eps}
\end{tabular}
\caption{\sf \small Evolution of the density of quintessence field. Left: for various values of energy 
fraction of dark matter decay remnants transformed into quintessence field: $10^{-16} $ (magenta), 
$5 \times 10^{-16}$ (cyan), $10^{-15}$ (blue), $5 \times 10^{-15}$ (green), $10^{-14}$ (red). The 
mass and self-coupling of the field quintessence are respectively $10^ {-6} $eV and $10^{-20} $; 
Center: for different values of quintessence mass $10^{-3} $eV (cyan), $10^{-5} $eV (magenta), 
$10^{-6} $eV (red) and $10^{-8} $eV (green); Right: for different values of self-coupling: $10^{-10} $ 
(cyan), $10^{-15} $, $10^{-20} $ and $10^{-25} $ (green). The difference between the density of 
quintessence field for the last 3 values of self-coupling is smaller than the resolution of this 
figure. The dash line is the observed value of the density of energy dark. $z_s$ is the redshift 
after which the density of condensate remains constant up to resolution of simulations. 
\label {fig:quinevol}}
\end{center}
\end{figure}

\subsubsection{Perturbations} Due to clustering of matter, an interaction in the dark sector can 
apriori induce a clustering in dark energy. However, there are stringent constraints on the 
anisotropic expansion of the Universe~\cite{anisoexpan} and clustering of dark 
energy~\cite{quinclust}. Therefore, it is necessary to verify that the model described above 
predicts an enough uniform distribution for the quintessence field to be consistent with data. 

After describing Einstein and Boltzmann equations for the scalar metric perturbations, linear 
perturbations of various matter components of this model, and few approximations to make an 
analytical calculation possible, we find the following relation between spatial fluctuation of 
quintessence field $\delta \phiq$ and velocity dispersion of dark matter $\delta {u_x}^i$:
\be
- V' (\bar{\phiq}, \bar {\rho}_x) {\partial}^i (\delta \phiq) = {\Gamma}_q \bar {\rho}_x 
\delta {u_x}^i \label {phidotueq}
\ee
This equation shows that the divergence of quintessence field fluctuations ${\partial}^i \delta \phiq$ 
follows the velocity dispersion of dark matter in opposite direction. However, its amplitude 
is largely reduced due to the very small decay width ${\Gamma}_q$. In addition, with the expansion of 
the Universe, $V' (\bar{\phiq}, \bar {\rho}_x)$ varies only very slightly, i.e. just the interaction 
between SDM and $\phiq$ changes. In contrast, $\bar {\rho}_x$ decreases by a factor of $a^{-3}(t)$ and 
even a gradual increase in the dark matter clumping and its velocity dispersion $\delta {u_x}^i$ is 
not enough to compensate the effect of its decreasing density~\cite {quinclust}. Therefore, we conclude 
that the spatial variation of $\phiq$ is very small and unobservant.

Finally, for testing this class of models against data, in addition to their impact on the expansion 
of the Universe and clustering that will be discussed in more detail in Sec. \ref{sec:deqft-param}, 
there are other means which can be used. In particular, a decaying heavy dark matter produces 
relativistic remnants which should be detectable directly if they are visible particles, or 
indirectly - through their effect on the evolution of large structure - if they are invisible. In 
fact some of recent observations prefer a larger number of relativistic species - usually described 
as the effective number of neutrinos~\cite{lssnuadd}, which apriori can be related to decay or 
interaction of dark matter. However, there are other explanations for these observations, for 
instance the existence of one or two sterile neutrinos~\cite{nushortbase}. Therefore, for the time 
being it is not possible to make a conclusion.

Alternatively, the lifetime of the decaying dark matter can be short. In this case, it has decayed 
longtime ago. Nonetheless, as we discussed before, the potential of quintessence field after its 
saturation stays constant and behave as a cosmological constant. This can explain the extreme 
fine-tuning of dark energy density with respect to dark matter in the early Universe. On the other 
hand, at late times the accelerating expansion of the Universe can destroy the coherence of 
quintessence condensate and dilute it. The study of this issue in the framework of quantum 
field theory is explained in the next section.

\section{Condensation of a quantum scalar field as the origin of dark energy} \subsectionmark {Condensation of a quantum scalar $\cdots$}\label{sec:deqft-cond} % ~10 pages
\subsection{Introduction} 
Quintessence models and many other phenomena in cosmology, particle physics, and condensed matter 
are associated to scalar fields which at large scales behave classically. Classical scalar fields 
have been first introduced in the fundamental physics in the framework of 
extensions to Einstein gravity~\cite{bransdicke,scalartensor}, and as a means for unification of 
gravity with other forces~\cite{dilaton,dilaton0}. They can be related to gravity models with 
conformal symmetry and its breaking which generates a scalar-tensor gravity~\cite{grconformal}. Thus, 
in these models the scalar field has a geometrical origin, at least at energy scales much smaller 
than Planck mass. By contrast, other known scalar fields such as Higgs in the Standard Model or 
Cooper pairs in condensed matter have a quantic point-like/particle nature, which has been 
experimentally demonstrated in both particle physics~\cite{higgsobs} and condensed 
matter~\cite{cooperpairobs}~\footnote{We should remind that this discrimination between particles 
and geometry loses its meaning in the geometric interpretation of fundamental models, specially in 
candidate models for quantum gravity such as string theory. However, at low energies differentiating 
between them may help better understand the underlying models.}. Their classical behaviour is 
associated to a special quantum state called a condensate - in analogy with condensation of droplets 
of liquid from particles or molecules of a vapor. In this section we briefly describe how a condensate 
can be formed at cosmological scales~\cite{p8,p49,p4}.

Decoherence of a scalar field due to its interaction with an environment, and the settlement of 
particles in one of the two minimum of a double-well potential is suggested as the origin for a 
nonzero vacuum energy and a prototype for landscape of string models~\cite{quinquantum,dedecohere}. 
This looks like counter intuitive because apriori the decoherence should reduce quantum correlations. 
In fact, it is exactly what happens. Rather than being in a quantum superposition of many energy 
states, by forming a condensate the energy distribution of particles is limited to one or few energy 
levels, see e.g.~\cite{multicond}. Although a condensate is a superposition state, it is more 
{\it deterministic}, i.e. has a smaller entropy, than the quantum state from which it is formed.

The formation of a condensate in cosmological environment raises several additional complexities,  
because the condensate must have an extension comparable to the size of the Universe. First of all, 
the mass of the scalar field and its coupling, both self and to other fields, must be very small. 
In the framework of the model explained in the previous section, quintessence particles are produced 
by the decay of a heavy particle. Consequently, at the moment of their production they must be highly 
relativistic. For instance, if they are produced during preheating after inflation or even at higher 
energy scales - for instance if they are associated to modulies in string theory - they must be 
initially relativistic. In laboratory, particles destined for condensation are usually cooled to 
very low temperatures before the process of condensation can occur. Due to their very weak coupling, 
quintessence particles cannot easily cool. These facts put stringent constraints on the condensation 
of a scalar field. Therefore, a comprehensive study is necessary to see whether a condensation can 
arise and to determine the necessary conditions for its occurrence. The first results of such an 
investigation are reported in~\cite{p8,p49,p4}.

In quantum mechanics, expectation values of hermitian operators associated to observables present the 
outcome of measurements. Therefore, it is natural to define the classical observable (component) of a  
quantum scalar field as its expectation value:
\be
\varphi (x) \equiv \langle \Psi|\Phi (x)|\Psi\rangle \neq 0 \label{classphi}
\ee
where $|\Psi\rangle$ is an element of the Fock space of the system. It is easily seen that a coherent 
state consisting of the superposition of indefinite number of particles in a single quantum state - 
presumably the ground state - behaves like a classical field as defined in (\ref {classphi}) i.e. 
$\langle \Psi|\Phi (x)|\Psi\rangle \neq 0$~\cite{condwave}. On the other hand, bosonic particles 
occupying the same energy state form a Bose-Einstein condensate. For this reason, the classical field 
$\varphi (x)$ is called a {\it condensate}. Using canonical representation, it is easy to see that 
for a limited number of free scalar particles $\langle \Psi|\Phi|\Psi \rangle = 0$. Nonetheless, in 
presence of an interaction, after renormalization a finite term can survive even when the state has 
a finite number of particles~\cite{infrenorm}. In this case, the field $\Phi$ can be considered to 
be {\it dressed}, which effectively presents an infinite number of virtual particles and satisfies 
the condensation condition~\cite{condint}. In such cases, the expectation value can be non-zero even 
on the vacuum. 

\begin{wrapfigure}{r}{0.5\textwidth}
\includegraphics[width=9cm]{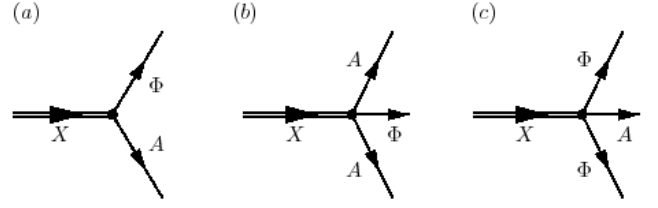}
\caption{\sf \small Decay models of heavy dark matter $X$. \label {decaymode}}
\end{wrapfigure}

\subsubsection{Formation of a quantum quintessence field} 
To investigate the reliability of quintessence models in a quantum field theoretical point of view, 
we consider a phenomenological model similar to one explained in Sec. \ref{sec:deqft-int}. The model 
consists of a heavy particle $X$ slowly decaying to two types of particles: a light scalar $\Phi$ and 
another field $A$ which can be an intermediate state or a collective notation for other fields. Here 
for simplicity it is assumed to be a scalar too. The particle $X$ and one of the remnants can be 
spinors, but the quintessence field $\Phi$ however must be a scalar. In the extreme density 
of the Universe after reheating, apriori the formation of Cooper-pair composite scalars from fermions 
is possible. However, this process needs a relatively strong interaction between fermions, and can 
arise in local phenomena such as Higgs mechanism and leptogenesis. But, due to very weak interaction 
of dark energy this does not seem to be plausible for $\Phi$. The simplest decay modes are shown in 
Fig. \ref{decaymode}.

Diagram (\ref{decaymode}-a) is the simplest decay/interaction mode. Diagram (\ref{decaymode}-b) is a 
prototype decay mode when $X$ and $\Phi$ share a conserved quantum number or $A$ and $\bar{A}$ (here 
$A=\bar{A}$ is considered) have a conserved quantum number. For instance, one of the favorite 
candidates for $X$ is a (s)neutrino decaying to a much lighter (pseudo-)scalar field and another 
(s)neutrino carrying the same leptonic number~\cite{rnu}. With seesaw mechanism between superpartners, 
a mass split between right and left (s)neutrinos, respectively corresponding to $X$ and $\Phi$, can 
occur. Depending on the conservation or violation of $R$-symmetry, the other remnant can be 
another scalar superpartner, Higgs or Higgsino.

The Lagrangian of this model is:
\bea
{\mathcal L} &=& {\mathcal L}_{\Phi} + {\mathcal L}_{X} + {\mathcal L}_{A} + {\mathcal L}_{int} 
\label{lagrangetot} \\
{\mathcal L}_{\Phi} &=& \int d^4 x \sqrt{-g} \biggl [\frac{1}{2} g^{\mu\nu}
{\partial}_{\mu}\Phi {\partial}_{\mu}\Phi - \frac{1}{2}m_{\Phi}^2 {\Phi}^2 - 
\frac{\lambda}{n}{\Phi}^n \biggr ] \label{lagrangphi} \\
{\mathcal L}_{X} &=& \int d^4 x \sqrt{-g} \biggl [\frac{1}{2} g^{\mu\nu}
{\partial}_{\mu}X {\partial}_{\mu}X - \frac{1}{2}m_X^2 X^2 \biggr ] 
\label{lagrangx}\\
{\mathcal L}_{A} &=& \int d^4 x \sqrt{-g} \biggl [\frac{1}{2} g^{\mu\nu}
{\partial}_{\mu}A {\partial}_{\mu}A - \frac{1}{2}m_A^2 A^2 - 
\frac{\lambda'}{n'}A^{n'} \biggr ] \label{lagranga} \\
{\mathcal L}_{int} &=& \int d^4 x \sqrt{-g} \begin{cases} 
\mg \Phi X A, & \text{For (\ref{decaymode})-a} \\
\mg \Phi X A^2, & \text{For (\ref{decaymode})-b} \\ 
\mg {\Phi}^2 XA, & \text{For (\ref{decaymode})-b} \end{cases}  
\label{lagrangint}
\eea
In the rest of this section we only describe the case (a) in detail. Note that no self-interaction is 
considered for $X$. The self-interaction of $A$ can be an effective description for interaction of 
fields collectively presented by $A$. Very weak interaction constraint on dark energy means that 
couplings $\lambda$ and $\mg$ must be very small. In a realistic particle physics model, 
renormalization and non-perturbative effects can lead to complicated potentials for scalar fields. 
An example relevant to dark energy is a pseudo-Nambu-Goldstone boson field with a shift 
symmetry~\cite{quinpng} which protects the very small mass of the quintessence field. The power-law 
potential considered in (\ref{lagrangphi}) can be interpreted as the dominant term or one of the 
terms in a potential with shift symmetry. 

\subsection{Decomposition and evolution equations} 
As we discussed above, our main aim is to study the evolution of quintessence condensate. Following 
definition (\ref{classphi}) we decompose $\Phi (x)$ to a condensate and a quantum component:
\be
\Phi (x) = \varphi (x) I + \phi (x) \quad \quad \langle \Phi \rangle \equiv 
\langle \Psi|\Phi|\Psi \rangle = \varphi (x) \quad \quad  \langle \phi \rangle 
\equiv \langle \Psi|\phi|\Psi \rangle = 0 \label{decomphi} 
\ee
where $I$ is the unit operator. We remind that in (\ref{decomphi}) both classical and quantum 
components depend on the spacetime $x$. We do not assume a homogeneous Universe, but only consider 
small anisotropies. We assume $\langle X \rangle = 0$ and $\langle A \rangle = 0$. The justification 
for these assumptions is the large mass and perturbative interactions of $X$ and $A$ which reduce 
their number and quantum effects. Later we show quantitatively that when the mass of a field is 
large, the minimum of effective potential of its condensate approaches zero. As $X$ and $A$ have a 
very weak interaction with $\Phi$, their evolution can be studied semi-classically by using the 
Boltzmann equation with a collisional term~\cite{p27,p25}. A more precise formulation should use the 
full Schwinger-Keldysh / Kadanoff-Baym formalism. This is a work in progress~\cite{houricondqm}.

After insertion of decomposition (\ref{decomphi}) into the Lagrangian (\ref{lagrangetot}), the 
evolution equation for the condensate $\varphi$ with interaction (a) is obtained by application of 
variational principle:
\be
\frac{1}{\sqrt{-g}}{\partial}_{\mu}(\sqrt{-g} g^{\mu\nu}{\partial}_{\nu}
\varphi) + m_{\Phi}^2 \varphi + \frac{\lambda}{n}\sum_{i=0}^{n-1} (i+1)
\binom{n}{i+1}{\varphi}^i\langle{\phi}^{n-i-1}\rangle - \mg \langle 
XA\rangle = 0 \quad \quad \text{for (\ref{decaymode}-a)} \label {dyneffa}
\ee
Note that in (\ref{lagrangint}) non-local interactions, i.e. terms containing derivatives of 
$\varphi$ do not contribute in the evolution of $\varphi$ because they are all proportional to 
$\phi$. After taking expectation value of the operators they cancel out because 
$\langle \phi \rangle = 0$ by definition. The expectation values depend on the quantum state of 
the system $|\Psi\rangle$ which presents the state of all particles in the system. We should 
remind that the mass of quantum component $\phi$ and thereby its evolution depends on $\varphi$. 
Moreover, through the interaction of $\Phi$ with $X$ and $A$ evolution of all the constituents 
of this model are coupled. In fact, for $n \geqslant 2$ the expectation values 
$\langle \phi^{(n-i-1)} \rangle~~ i=0,\cdots,n-1$ modify the mass and self-coupling of $\varphi$. 
Another important observation is that in general, the effective potential of condensate $\varphi$ 
is not the same as the classical potential in the Lagrangian.

\begin{wrapfigure}{r}{0.5\textwidth}
\vspace{-1cm}
\bea
&& \hspace{-0.5cm}\includegraphics[width=8cm]{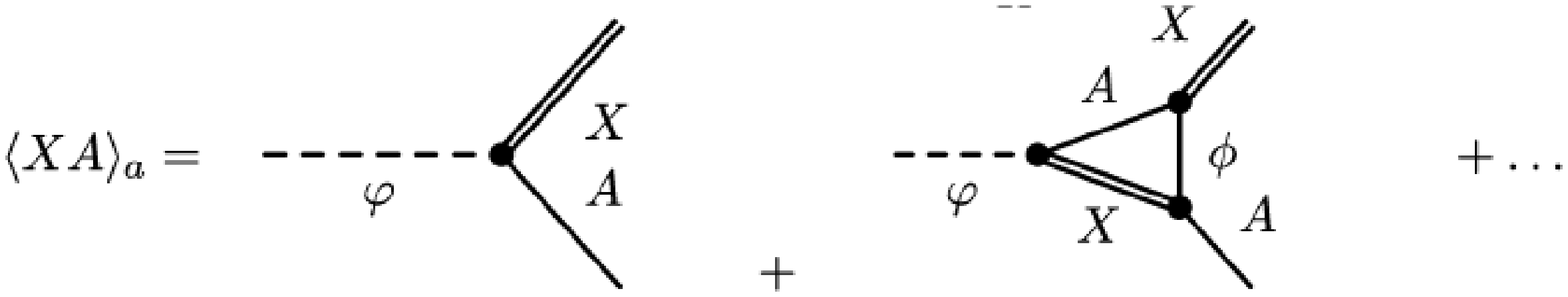} \label {xathreediag} \\
&& \hspace{-0.5cm}\includegraphics[width=8cm]{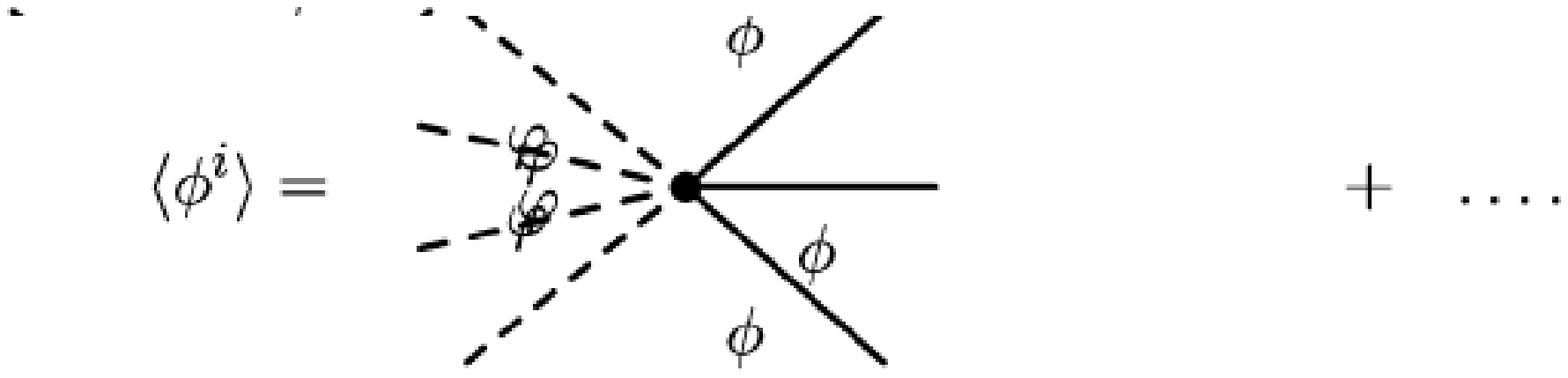} \label {phiphidiag} \\
&& \hspace{-0.5cm}\includegraphics[width=8cm]{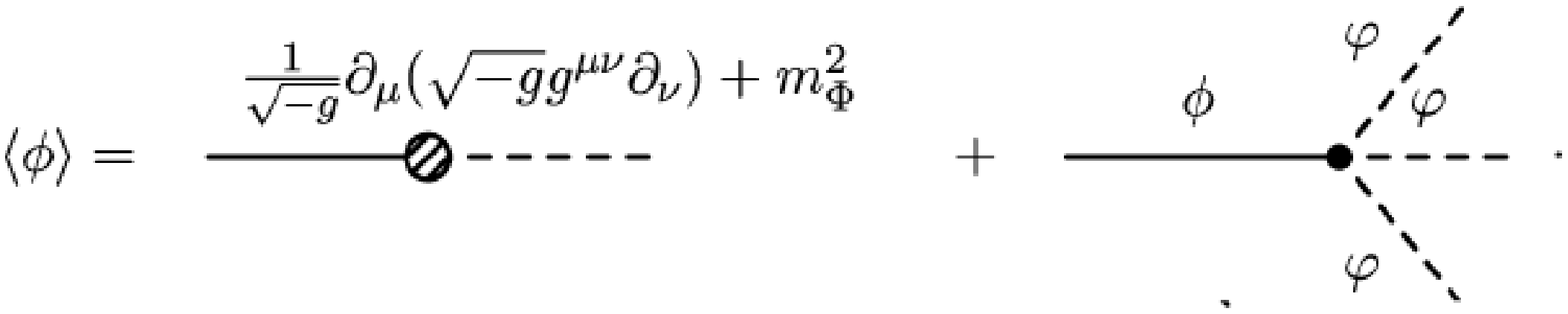} \label {phiphivardiag}
\eea
\caption{\sf \small Diagrams included in the calculation of expectation values for decay model (a). 
The dash line presents the condensate component $\varphi$. \label{fig:inindiag}}
\end {wrapfigure}

We use Schwinger-Keldysh closed time path integral formalism to calculate expectation values in 
(\ref{dyneffa}) at zero-order (tree diagrams). The next relevant diagram is of order $g^3$, see 
Fig. \ref{fig:inindiag}. Therefore, under the assumption of the small couplings for the quintessence 
field, higher order diagrams are negligible. The decomposition of $\Phi$ also affects the 
renormalization of the model. This issue has been already studied~\cite{infrenorm} and we do not 
consider it here. We simply assume that masses and couplings in the model have their values after 
renormalization. One reason for overlooking this issue is the fact that it should be 
discussed in the context of a full particle physics model. In accordance with the decomposition 
(\ref{decomphi}), the sum of graphs in (\ref{phiphivardiag}) is null because they correspond to 
the field equation of $\varphi$ (\ref{dyneffa}). Finally, the expectation values at zero order are:
\bea
\langle XA\rangle_a & = & -i\mg \int \sqrt{-g} d^4y \varphi (y)
\biggl [G_A^> (x,y) G_X^> (x,y) - G_A^< (x,y) G_X^< (x,y)\biggr ] 
\label {avalxa} \\
\langle \phi^i\rangle & = & -i\lambda \int \sqrt{-g} d^4y \varphi^{n-i} (y)
\biggl [[G_{\phi}^> (x,y)]^i - [G_{\phi}^< (x,y)]^i \biggr ] \label {valphiphi}
\eea
where $G^>$ and $G^<$ are respectively advanced and retarded propagators, related to the Feynman 
propagator $G_F (x,y) = G^> (x,y) \Theta (x^0-y^0) + G^< (x,y) \Theta (y^0-x^0)$ which can be 
determined by solving following field equations:
\bea
&&\frac{1}{\sqrt{-g}}{\partial}_{\mu}(\sqrt{-g} g^{\mu\nu}{\partial}_{\nu}
G^{\phi}_F (x-y)) + (m_{\Phi}^2 + (n-1) \lambda {\varphi}^{n-2}) 
G^{\phi}_F (x-y) = -i \frac{{\delta}^4 (x-y)}{\sqrt{-g}} \label {propagphi} \\
&&\frac{1}{\sqrt{-g}}{\partial}_{\mu}(\sqrt{-g} g^{\mu\nu}{\partial}_{\nu}G^i_F 
(x-y)) + m_i^2 G^i_F (x-y) = -i\frac{{\delta}^4 (x-y)}{\sqrt{-g}} \quad , \quad
i = X, A \label {propagx}
\eea
It is remarkable that even at zero (classical) order $G^{\phi}_F (x-y)$ depends on the condensate field 
$\varphi$. The coupling between quantum component $\phi$ and the condensate $\varphi$ is the origin of 
the back-reaction of the condensate formation on the quantum fields. Assuming that $\Phi$ particles 
are produced only through the decay of $X$, the initial value of $\varphi = 0$ and the coupling 
between $\phi$ and $\varphi$ is very small. With the growth of the $\varphi$ amplitude, the 
effective mass of $\phi$ particles increases. In turn, this affects the growth of the condensate 
because due to an energy barrier $\phi$ particles will not be able to join the condensate anymore. 
This negative feedback prevents an explosive formation of the condensate. We should remind that for 
a full consideration of the backreaction of interactions we must consider propagators that include 
2-Particle Irreducible (2-PI) self-energy correction. However, their inclusion makes the problem 
completely unsolvable analytically. For this reason the full formulation is left for a future 
work~\cite{houricondqm} that will study this model through numerical simulations. 

We expect a rapid decoherence of $\phi$ particles and other species due to the fast acceleration of 
the Universe. It pushes long wavelength fluctuations (IR modes) out of the horizon. In turn, they 
play the role of an environment for the decoherence of shorter modes. Thus, $X$, $A$, and $\phi$ 
particles can be considered as semi-classical and the evolution of their density is ruled by 
classical Boltzmann equations, see e.g.~\cite{p27}. A complete treatment of the model as a 
non-equilibrium quantum process must include Kadanoff-Baym equations. It would be necessary when 
this model is studied in the context of a realistic particle physics.

The last set of equations to be considered for a consistent and full solution of the model are Einstein 
equations that determine the relation between the geometry of the spacetime and the evolution of its 
matter content. Apriori it is important to consider the backreaction of matter and dark energy 
anisotropies on the metric, at least in linear order which is a good approximation at large scales even 
today. However, due to the complexity of the model, we only consider a homogeneous background metric. 
It is shown in~\cite{p8} that at linear order propagators in a background with small scalar 
fluctuations are simply $G_h(x,y)(1+\psi)$ where $\psi$ is the gravitational potential in the Newtonian 
gauge and $G_h(x,y)$ is the propagator in a homogeneous background. This relation can be used to 
estimate the effect of metric fluctuations on the evolution of quintessence condensate. Under these 
approximations only the evolution of expansion factor $a(t)$, i.e. the Friedmann equation has special 
importance for the determination of condensate evolution.

During radiation domination epoch the density of non-relativistic particles such as $X$ is by 
definition negligible, and evolution of $a (t)$ is governed by relativistic species which are not 
considered here explicitly. From the observed density of dark energy we can conclude that in this 
epoch its density was much smaller than other components, and had negligible effect on the evolution 
of expansion factor. In the matter domination epoch both $X$ and $A$ are assumed to be 
non-relativistic. If the lifetime of $X$ is much shorter than the age of the Universe at the 
beginning of matter domination epoch, most of $X$ particles have decayed, and $X$ does not play a 
significant role in the evolution of $a (t)$ which is determined by other non-relativistic species. 
If the lifetime of $X$ is much larger than the age of the Universe, then $X$ particles can have a 
significant contribution in the total density of matter. Considering the very slow decay of $X$, in 
the calculation of $a (t)$, at lowest order it can be approximately treated as stable and $a (t)$ 
evolves similar to the case of a CDM model. A better estimation of $a (t)$ can be obtained by taking 
into account the decay of $X$ to relativistic particles~\cite{p29}. Once again here we use the 
simplest approximation because the problem in hand is very complicated and we want to keep the 
evolution of $a (t)$ decoupled from other equations such that we can obtain an analytical 
approximation. At late times when the density of the condensate becomes comparable to matter density, 
the full theory including Boltzmann equations must be solved. In this case the evolution of $a (t)$ 
is closely related to the evolution of the quintessence condensate and a full numerical solution is 
necessary.

\subsection{Quantum state} 
Propagators and expectation values described in the previous section are defined for the quantum state 
of constituents of the Universe content. Therefore, before any attempt to calculate these quantities we 
must know their quantum state\footnote{In Heisenberg picture states are constant and operators 
vary with time. Therefore one can calculate all quantities for vacuum. Nonetheless, the state is 
necessary deciding the initial conditions necessary for the solution of propagators and condensate 
equations, and interpretation of observed phenomena because we usually associate an observation to the 
observed system rather than an abstract operator.}.

Due to weak interactions between particles in this model, after their decoherence they can be 
considered as freely-scattering particles, and therefore their quantum state $|\Psi_f\rangle$ can be 
approximated by direct multiplication of single particle states:
\be
|\Psi_f\rangle = \sum_{p_j} \bigotimes_{i,j} f^i (x, \{p_j\}, \varphi)|p^i_j\rangle  \label{psif}
\ee
The indices $i$ and $j$ respectively present the species type and particle number, and $\{p_j\}$ the 
momentum of all states. Distributions $f^i (x, \{p_j\}, \varphi)$ can be related to quantum properties 
of the system by using Wigner function~\cite{wigner}. Note that due to the dependence of masses and 
couplings on the condensate $\varphi$, distribution of semi-classical particles depend on this 
quantity too. By projecting $\Psi$ into the coordinate space we can express $|\Psi|^2$ as a functional 
of Wigner function:
\be
|\Psi|^2 = \Psi^*(x) \Psi(y) = \Psi^* (\bar{x}+\frac{X}{2}) \Psi (\bar{x}-\frac{X}{2}) = 
\frac{\sqrt {-g}}{(2\pi)^4} \int d^4 p P(p, \bar{x}) e^{-ip.X} \quad , \quad \bar{x} \equiv 
\frac {x+y}{2} \quad , \quad X \equiv x-y \label{wignerfunc}
\ee
In the classical limit Wigner function $P(p, \bar{x})$ approaches the classical distribution 
function $f (p, \bar {x})$ which can be determined in a consistent way from classical Boltzmann 
equations or their quantum extensions Kadanoff-Baym equations. In fact, it has been 
shown~\cite{kadanofbaymstoch} that distributions can be directly related to Green's functions:
\bea
&& \langle \hat{N} (\vec{k},t) \rangle \equiv \omega_k f (k,x, \varphi) = {\mathcal D} \langle 
\phi (-k,t)\phi (k,t')\rangle \biggl |_{t=t'} \label{greentodist} \\
&& {\mathcal D} \equiv \frac{1}{2}\omega_k + \frac{1}{\omega_k}\frac{\partial}{\partial t}
\frac{\partial}{\partial t'} + i (\frac{\partial}{\partial t} - \frac{\partial}{\partial t'}) 
\label{timediffop}
\eea
where $\hat{N}$ is number operator. As mentioned earlier, in the study performed in~\cite{p8} 
Boltzmann equations are not solved along with the evolution equation of the condensate and a thermal 
approximation was used in place.

Determination of the quantum state of the condensate is less straightforward and no general expression 
or a procedure to obtain it is available. Nonetheless, it is easy to verify that Glauber's coherent 
states~\cite {cohereglauber} satisfy the condition (\ref{classphi}) for 
condensates~\cite{condcoherestate}. After decomposition of quintessence field to creation and 
annihilation operators:
\be
\Upsilon \equiv a (\eta)\Phi (x) = \sum_k [\um_k (x) a_k + \um_k^* (x) a^{\dagger}_k] \quad , 
\quad [a_k, a^{\dagger}_{k'}] = \delta_{kk'} \quad [a_k,a_{k'}] = 0 \quad 
[a^{\dagger}_k,a^{\dagger}_{k'}] = 0\label{canon}
\ee
where $\um_k (x) \equiv \um_k (\eta)e^{-i\vec{k}.\vec{x}} $ is a solution of the free field equation, a 
coherent state is defined as:
\be
|\Psi_C\rangle \equiv e^{-|C|^2} e^{C a_0^{\dagger}} |0\rangle = e^{-|C|^2} \sum_{i=0}^\infty 
\frac {C^i(x)}{i!}(a_0^{\dagger})^i |0\rangle \label{condwavef}
%\quad , \quad C_{\chi} (t) = \frac {1}{\sqrt{2 m_\Upsilon H^3}} \chi (t) \cos (m_\Upsilon t) 
\ee
It can be verified that this state satisfies the relation~\cite{condwave}:
\be
a_0 |\Psi_C\rangle = C |\Psi_C\rangle \label{condcond}
\ee
From decomposition of $\phi$ to creation and annihilation operators (\ref{canon}) we find:
\be
\chi (x) \equiv a \langle \Psi_C |\Phi|\Psi_C\rangle = 
C \um_0 (x) + C^* \um_0^* (x)\label{condexp}
\ee
Here we have adapted the original formula of~\cite{condwave} for a homogeneous FLRW cosmology. As 
$\chi$ is a real field the argument of $C$ is arbitrary, and therefore we assume that $C$ is real:
\be
C = \frac{\um_0 (x) + \um_0^* (x)}{\chi (x)} \label{condc}
\ee 
%The physical interpretation of this state is that any number of particles with zero momentum can be added to a condensate, i.e. their chemical potential is null. Nonetheless, a condensate can be formed only in an interacting system, and in general the chemical potential of such systems is not zero. Moreover, 
Condensates produced in laboratory usually include multiple energy levels with 
approximately decoupled condensates at each energy level, see e.g.~\cite{multicondlab}. For these more 
general cases the definition of a condensate can be generalized in the following manner: Consider a 
system with a large number of scalar particles of the same type. Their only discriminating observable 
is their momentum. The distribution of momentum is discrete if the system is put in a finite volume. 
Such setup contains sub-systems similar to (\ref{condwavef}) consisting of particles with momentum 
$\vec{k}$:
\be
|\Psi_k\rangle \equiv A_k e^{C_k a_k^{\dagger}} |0\rangle = A_k \sum_{i=0}^N 
\frac {C_k^i}{i!}(a_k^{\dagger})^i |0\rangle \label{condwavek}
\ee
where $A_k$ is a normalization constant. It is easy to verify that this state satisfies the relation:
\be
a_k |\Psi_k\rangle_N = C_k |\Psi_k\rangle_{(N-1)} \label{condcondk}
\ee
If $N \rightarrow \infty$, the identity (\ref{condcondk}) becomes similar to (\ref{condcond}) and the 
expectation value of the scalar field on this state is non-zero. Therefore, we define a 
multi-condensate or generalized condensate state as a state in which every particle belongs to a 
sub-state of the form (\ref{condwavek}): 
\bea
&& |\Psi_{GC}\rangle \equiv \sum_k A_k e^{C_k a_k^{\dagger}} |0\rangle = \sum_k A_k 
\sum_{i=0}^{N \rightarrow \infty} \frac {C_k^i}{i!}(a_k^{\dagger})^i |0\rangle 
\label{condwaveg} \\
%&& C_{\chi_k} (t) = \frac {1}{\sqrt{2 E_k H^3}} \chi_k (t) \cos (E_k t) \quad , \quad E_k^2 \equiv k^2 + m_\Upsilon^2 \label{condcondg} \\
&& \chi (x,\eta) \equiv a (\eta) \langle \Psi_{GC}|\Phi|\Psi_{GC}\rangle = \sum_k  
C_k \um_k (x) + C^*_k \um_k^* (x) \label{condexpg}
\eea
The state $|\Psi_{GC}\rangle$ satisfies the equality (\ref{condcondk}). Coefficients $A_k$ 
determine the relative amplitudes of condensate at each momentum with respect to each others. Using 
(\ref{condexpg}), the evolution equation of the field determines how $C_k$'s evolve. It is easy to 
verify that the energy density and effective number density of $|\Psi_{GC}\rangle$, respectively 
defined as the expectation value of $m_\Phi^2 \Phi^2/2$ and the number operator 
$\sum_k a_k^{\dagger} a_k$, are finite:
\bea
&& \langle \Psi_{GC} |\frac{m_\Phi^2 \Phi^2 }{2} |\Psi_{GC}\rangle = m^2 a^{-2} (\eta) \sum_k 
\biggl [{\mathcal Re} (C_k^2 \um_k (x)) + |\um_k (x) C_k|^2 + \frac {1}{2} \biggr ] 
\label{condener} \\
&& \langle \Psi_{GC} |\sum_k a_k^{\dagger} a_k|\Psi_{GC}\rangle = \sum_k |C_k|^2 
\label{condnum}
\eea
The reason for the finiteness of these quantities despite the presence of infinite number of 
states in (\ref{condwaveg}) is the exponentially small amplitude of the components with 
$N \rightarrow \infty$.

When we calculate the propagators of $\phi$ we should take into account the contribution of all 
$\Phi$ particles in the wave function of $\Psi$, including the condensate. Therefore:
\be
|\Psi^{(\Phi)}|^2 \approx f^{(\Phi)} (p, \bar{x}) + f^{(\varphi)}(\bar{x}) \label{psiphi}
\ee
where $f^{(\varphi)}$ is the contribution of the condensate and $f^{(\Phi)}$ the distribution of 
non-condensate decohered particles which can be treated classically. Note that the separation of two 
components in (\ref{psiphi}) is an approximation and ignores the quantum interference between 
{\it free} $\Phi$ particles and the condensate. This approximation is valid if the self-interaction of 
$\Phi$ is weak and the non-condensate component decohere rapidly. The advantage of the generalized 
coherent state explained above for dark energy is the fact that quintessence particles do not need 
to lose completely their energy to join the condensate. This significantly softens the constraint 
imposed by the tiny interaction of $\Phi$ on the formation of a condensate. We should remind that 
many other coherent states, e.g. for special geometries exist in the literature~\cite{qmcohere}.

\subsection{Solution of evolution equation of quintessence condensate} 
When interactions are neglected and after field redefinition $\chi \equiv a\varphi$ and taking the 
Fourier transform with respect to spatial coordinates the field equation to solve takes the following 
form:
\be
\um_k'' + k^2 \um_k + (a^2 m^2 - \frac{a''}{a}) \um_k = \begin{cases} 0 & 
\text{For evolution of condensate} \\ 
-i \frac {\delta (\eta - \eta')}{a} & \text{For propagators} \label{prophomo} \end {cases}
\ee
where $\eta$ is the conformal time. After adding the contribution of a non-vacuum state, the Feynman 
propagator $G (\eta,\eta')$ has the following expansion:
\be
iG_k (\eta, \eta') = \biggl [{\mathcal A}^>_k \um_k (\eta)\um^*_k (\eta') + {\mathcal B}^>_k 
\um^*_k (\eta) \um_k (\eta')\biggr ] \Theta (\eta - \eta') + \biggl [{\mathcal A}^<_k 
\um_k (\eta)\um^*_k (\eta') + {\mathcal B}^<_k \um^*_k (\eta) \um_k (\eta')\biggr ] 
\Theta (\eta' - \eta) \label{progexpand}
\ee
where ${\mathcal A}^>_k$, ${\mathcal B}^>_k$, ${\mathcal A}^<_k$ and ${\mathcal B}^<_k$ are integration 
constants. For free propagators on non-vacuum states, it is possible to include the contribution of 
the state in the boundary conditions imposed on the propagator, see Appendix-A in~\cite{p8}. 
This leads to following relations between integration constants and the wave function of the system:
\bea
&&{\mathcal A}^>_k = 1 + {\mathcal B}^>_k \quad , \quad {\mathcal B}^<_k = 
1 + {\mathcal A}^<_k \quad , \quad {\mathcal A}^<_k = {\mathcal B}^>_k = \sum_i \sum_{k_1 k_2 
\ldots k_n} \delta_{kk_i} |\Psi_{k_1 k_2 \ldots k_n}|^2 \label{abpsi} \\
&& G_k^> (\eta,\eta')\biggl |_{\eta = \eta'} = G_k^< (\eta,\eta')\biggl |_{\eta = 
\eta'} \label {consistcond} \\
&& \um_k^{'} (\eta) \um^*_k (\eta) - \um_k (\eta) \um^{'*}_k (\eta)= \frac {-i}{a(\eta)} 
\label{derivcond}
\eea

\subsubsection{Initial conditions for propagators} Field equations are second order differential 
equations and a complete description of the solutions needs the initial value of the field and its 
derivative. Alternatively, they can be treated as a boundary value problem in which the values of the 
field at two different epochs are constrained. A physically motivated initial condition for a bounded 
system, including both Neumann and Dirichlet conditions is~\cite{qftinit}:
\be
a^{-1}\partial_{\eta} \um_k = -i {\mathcal K} \um_k \label{geninitorth} 
\ee
where the spacelike vector $n^{\mu}$ is normal to the boundary and defined as 
$n^{\mu} = a^{-1}(1,0,0,0)$, and $\um$ is a solution of the differential equation. The constant 
${\mathcal K}$ depends on the scale $k$. In a cosmological setup, the initial condition constraint 
(\ref{geninitorth}) must be applied to both past (initial) and future (final) boundary 
surfaces~\cite{qftinit}. In the case of propagators, they are applied only to one of the past or 
future limit, respectively for advanced and retarded propagators. In each case the other boundary 
condition is replaced by consistency condition (\ref{consistcond}). Assuming different values for 
${\mathcal K}$ on these boundaries, we find:
\bea
&& {\mathcal K}_j = i\frac {\um^{'}_k (\eta_j)}{a_j \um_k(\eta_j)}, \quad j = i, f \label {alphavac} \\
&& |\um_k (\eta_j)|^2 = \frac {1}{a^2(\eta_j) ({\mathcal K}_j (k,\eta_j) + 
{\mathcal K}^*_j (k,\eta_j))}, \quad |\um^{'}_k (\eta_j)|^2 = 
\frac {|{\mathcal K}_j (k,\eta_j)|^2}{{\mathcal K}_j (k,\eta_j) + 
{\mathcal K}^*_j (k,\eta_j)} \label{uband}
\eea
In cosmological context, ${\mathcal K}_f$ can be fixed based on observations, but ${\mathcal K}_i$ 
is unknown and leaves one model-dependent constant that should be fixed by the physics of the early 
Universe. This arbitrariness of the general solution or in other words the vacuum of the theory is 
well known~\cite{vacuu}. In the case of inflation - in De Sitter spacetime - a class of possible 
vacuum solutions called $\alpha$-vacuum allow the following expression for ${\mathcal K}$:
\be 
{\mathcal K}_i , {\mathcal K}_f = \sqrt {k^2/a^2_{i,f} + m^2} 
\label{bunchdavies}
\ee
and one obtains the well known Bunch-Davies solutions~\cite{qftinit}. We use this choice for the 
quintessence model studied here. 

\subsubsection{Radiation domination era} The $X$ particles are presumably produced during reheating 
epoch~\cite{massrenor,wimpzilla} and their decay begins afterward. In this epoch relativistic 
particles dominate the energy density of the Universe. Therefore, the expansion factor $a (\eta)$ 
can be determined independently. Fortunately, in this epoch homogeneous field equations have exact 
and well known solutions~\cite{qmcurve}, and after applying the WKB approximation the full solutions 
of the evolution equation of quintessence condensate including interactions can be obtained:
\bea
U_k &\approx& \sqrt{\frac{\eta_0}{\eta}} \exp \biggl (\frac{1}{2} \sum_{\alpha,\beta}
B'_{\alpha\beta} \sin~(2\alpha \ln \frac{\eta}{\eta_0} - \frac {\beta \eta^2}
{4 \eta_0^2}) \biggl [\frac{\eta}{\eta_0} + A'_{\alpha\beta} \cos~(2\alpha \ln 
\frac{\eta}{\eta_0} - \frac {\beta \eta^2}{4 \eta_0^2}) \biggr ] \biggr ) \times \nonumber \\
&& \exp \biggl (-\frac{i}{4} \sum_{\alpha,\beta} \biggl \{ \biggl [\frac{\eta}{\eta_0} + 
A'_{\alpha\beta} \cos~(2\alpha \ln \frac{\eta}{\eta_0} - \frac {\beta \eta^2}{4 \eta_0^2}) 
\biggr ]^2 - {B'}^2_{\alpha\beta} \sin^2 (2\alpha \ln \frac{\eta}{\eta_0} - 
\frac {\beta \eta^2}{4 \eta_0^2}) \biggr \} \biggr ) \label{uapprox} \\
V_k - iU_k &\approx& \sqrt{\frac{\eta_0}{\eta}} \exp \biggl (-\frac{1}{2} 
\sum_{\alpha,\beta} B'_{\alpha\beta} \sin~(2\alpha \ln \frac{\eta}{\eta_0} - 
\frac {\beta \eta^2}{4 \eta_0^2}) \biggl [\frac{\eta}{\eta_0} + A'_{\alpha\beta} 
\cos~(2\alpha \ln \frac{\eta}{\eta_0} - \frac {\beta \eta^2}{4 \eta_0^2}) \biggr ] 
\biggr ) \times \nonumber \\
&& \exp \biggl (\frac{i}{4} \sum_{\alpha,\beta} \biggl \{ \biggl [\frac{\eta}{\eta_0} + 
A'_{\alpha\beta} \cos~(2\alpha \ln \frac{\eta}{\eta_0} - \frac {\beta \eta^2}{4 \eta_0^2}) 
\biggr ]^2 - {B'}^2_{\alpha\beta} \sin^2 (2\alpha \ln \frac{\eta}{\eta_0} - 
\frac {\beta \eta^2}{4 \eta_0^2}) \biggr \} \biggr ) \label{vapprox}
\eea
where $\eta_0$ is the initial conformal time, and $A'$ and $B'$ are constants depending on the 
parameters of the model $\theta_i \equiv \sqrt{2 a_0 \eta_0 m_i} = \sqrt{\frac{2m_i}{H_0}},~ \alpha_i 
\equiv \frac{k^2\eta_0}{2 a_0 m_i} = \frac{k^2 H_0\eta_0^2}{2m_i},~i = \Phi,~\cx,~\ca$ and 
$\alpha = (\pm \alpha_A \pm \alpha_X)$ and $\beta = (\pm \theta_A^2 \pm 
\theta_X^2)$. The presence of a real exponential term in both independent 
solutions of the evolution equation, and the phase difference between them means that in the radiation 
domination epoch there is always a growing term that assures the accumulation of the condensate. 
However, due to the smallness of coefficients $A'_{\alpha\beta}$ which are proportional to $\mg^2$, 
the growth of the condensate can be very slow. Therefore, we conclude that in this regime the 
production of $\Phi$ particles by the slow decay of $X$ is enough to produce a quintessence 
condensate. Figure \ref{fig:uv} shows $V_k$ and $U_k$ for a choice of parameters. We should remind 
the similarity of these solutions to parametric resonance during preheating~\cite{preheat}. This is 
not a surprise because the form of the evolution equations of these models are very similar.
\begin{figure}[h]
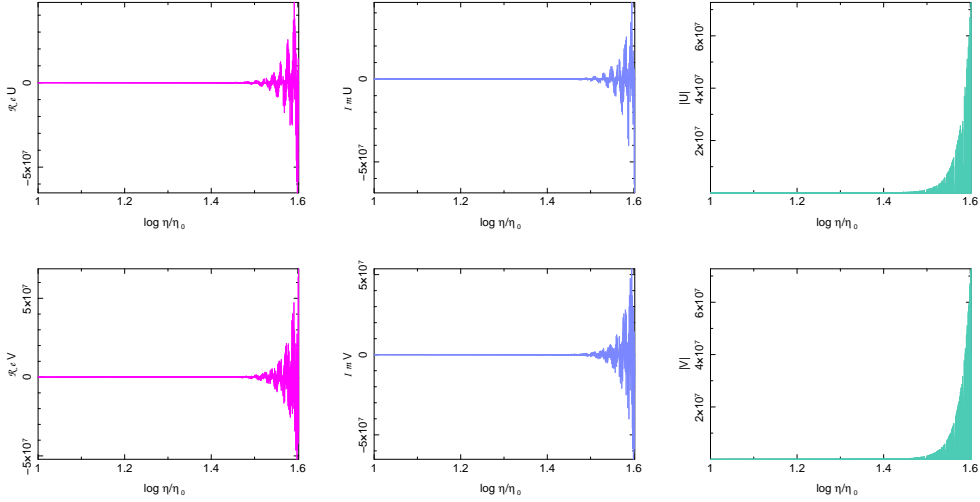

\begin{center}
\begin {tabular}{ccc}
\includegraphics[height=4cm,angle=-90]{reu.ps} & 
\includegraphics[height=4cm,angle=-90]{imu.ps} &
\includegraphics[height=4cm,angle=-90]{absu.ps} \\
\includegraphics[height=4cm,angle=-90]{rev.ps} & 
\includegraphics[height=4cm,angle=-90]{imv.ps} &
\includegraphics[height=4cm,angle=-90]{absv.ps} 
\end {tabular}
\end {center}
\caption{\sf \small Real, imaginary, and absolute value of $U_k$ and $V_k$ for $\alpha = 0$ and 
$\beta = 100$. The general aspects of these functions are not very sensitive to $\alpha$ and are very 
similar for $\beta \gtrsim 10$. Note that although there are resonant jumps in the value of $U$ and 
$V$, due to complicated interaction terms they are not regular as in the preheating.
\label{fig:uv}}
\end{figure}
\subsubsection{Backreaction} An exponential growth of the condensate for ever would be evidently 
catastrophic for this model. We show below that during matter domination era the faster expansion of 
the Universe stops the growth. Moreover, if $X$ particles have a short lifetime and decay completely 
before the end of the radiation domination epoch, production term in (\ref{avalxa}) becomes negligibly 
small. Due to many simplifying assumptions we had to make to be able to obtain approximative analytical 
solutions (\ref{uapprox}) and (\ref{uapprox}), some other issues must be also taken into account. 
For instance, we neglected the effect of free $\phi$ particles. Energy transfer between them and 
the condensate can lead to evaporation of the latter. This effect would be consistently taken into 
account if we solve Boltzmann/Kadanoff-Baym equations along with evolution equation of the condensate 
and add 2PI terms in the evolution of propagators. 

\subsubsection{Matter domination era} In the matter domination epoch the relation between comoving and 
conformal time deviates from the previous era and consequently the evolution equation of fields is 
different and has the following form:
\be
\um_k'' + (k^2 + \frac{m^2 a_0^2\eta^4}{\eta_0^4} - \frac{2}{\eta^2}) \um_k = 0 
\label{progmatter}
\ee
In contrast to radiation domination epoch, this equation does not have known analytical solution. Only 
for two special cases of $m = 0$ and $k^2 = 0$ exact analytical solutions exist. Therefore, we have 
to use one of them, in preference the solution of $k^2 = 0$ which is closer to the case we are 
interested in, along with WKB approximation. When interactions are ignored the solution of the 
evolution equation has the following approximate expression:
\bea
\chi_k (\eta) & \xrightarrow[\lambda = 0]{\frac{\eta}{\eta_0} \gg 1} & \sqrt {\frac{2}
{\pi \beta'_\Phi}} \frac{\eta_0}{\eta} \biggl (1 - \frac{3k^2\eta_0}{2m_\Phi^2 \eta} + 
\ym_k (\eta) \biggr ) \biggl \{{c'}_k^{(a)} \sin \biggl (\beta' \frac{\eta^3}{\eta_0^3} 
(1- \frac{3k^2\eta_0}{2m_\Phi^2 \eta}) + \ym_k (\eta)\biggr ) + \nonumber \\
&& {d'}_k^{(a)} \cos \biggl (\beta' \frac{\eta^3}{\eta_0^3} (1 - \frac{3k^2\eta_0}
{2m_\Phi^2 \eta}) + \ym_k (\eta) \biggr )\biggr \} \label{chimatapp} \\
\ym_k (\eta) & = & \frac{i \mg^2}{4 (2\pi)^3 \pi \sqrt{\beta'_A \beta'_X}} 
\biggl \{\sum_{\alpha} C_\alpha (k, \bar{x}) \gamma (-2, i\alpha \frac{\eta^3}{\eta_0^3}) + 
\nonumber \\
&& \sum_{\alpha} C'_\alpha (k, \bar{x})\gamma (-\frac{1}{3}, i\alpha \frac{\eta^3}{\eta_0^3}) 
\biggr \} \label{vmsol}
\eea
where $C_\alpha$ and $C'_\alpha$ are proportional to the distributions of $A$ and $X$ particles and 
$\beta' \equiv \frac{a_0\eta_0 m}{3} = \frac{2m}{3H_0}$. At late times $\gamma$-functions in 
(\ref{vmsol}) approach a constant and $\bar{x}$ dependent terms i.e terms containing $f^{(A)}$ and 
$f^{(X)}$ decay very rapidly, as $(\eta_0/\eta)^6$ for terms containing one $f$, and $\chi_k (\eta)$ 
becomes an oscillating function that its amplitude decreases as $\eta_0/\eta$ with time. 
Consequently, $\varphi_k$ decreases as $\eta^3_0/\eta^3$ and the production of $\Phi$ in the decay 
of $X$ alone is not enough to compensate the expansion of the Universe, leading to a decreasing 
density of the condensate. Evidently, the validity of this conclusion depends on the precision of 
approximations considered in this calculation. In fact it is shown~\cite{p8} that linearized 
equations always arrive to the same conclusion, even when all interactions are taken into account.

To perform a nonlinear analysis of the condensate equation with self-interaction we first neglect 
quantum corrections. This means that we only consider the classical interaction term in (\ref{dyneffa}) 
for which the minimum of the potential is at origin. If for simplicity we neglect also the production 
term, the evolution equation becomes:
\be
{\chi}'' + (k^2 + a^2 m_{\Phi}^2 - \frac{2}{\eta^2}) \chi + \lambda a^{4-n} \chi^{n-1}(x) = 0
\label{aintclass}
\ee 
Using a difference approximation for derivatives but without linearization, we find that although at 
the beginning $\chi$ can grow irrespective of initial conditions, at late times it approaches zero. 
This means that this equation lacks a tracking solution. Another way of checking the absence of a 
tracking solution is the application of the criterion $\Gamma \equiv V"V/V'2 > 1$ proved to be the 
necessary condition for the existence of such solutions~\cite{trackingcond}. For equation 
(\ref{aintclass}) $\Gamma = n (n-1) / n^2 < 1$ for $n > 0$. This is a well known result. As mentioned 
in section \ref{sec:deqft-int}, only inverse power-law and inverse exponential potentials have a 
late time tracking solution~\cite{quin}.

When quantum corrections are added, the evolution equation depends on the coefficient $C$ which appears 
in the expression of propagators and determines the amplitude of the quantum state of the condensate. 
This coefficient depends inversely on $\chi$, see (\ref{condc}), and thereby induces a backreaction 
from the formation of the condensate to the propagators of $\phi$ and vis-versa. After adding these 
non-linear terms to the evolution equation of the quintessence condensate, it has the following 
approximate expression:
\bea
&& {\chi}'' + (k^2 + a^2 m_{\Phi}^2 - \frac{2}{\eta^2}) \chi + \frac{i}{3}\lambda^2 
a^{4-n} (\frac{2}{\pi \beta'_\Phi})^{n-2} e^{i\frac{(8-n)\pi}{6}} (\frac{\eta_0}{\eta})^{n-1} 
\sum_{\alpha,\beta} \beta^{-\frac{8-n}{3}} \gamma (\frac{8-n}{3}, -i\beta 
\frac{\eta^3}{\eta_0^3})~ e^{i (\alpha + \beta) \frac{\eta^3}{\eta_0^3}} 
\nonumber \\
&& \quad \quad \times \sum_{i = 1}^{n-1} \binom{n-1}{i} (\frac{2}{\pi\beta'_\Phi})^{n-i} 
\cos^{2(n-i)}(\beta'_\Phi \frac{\eta^3}{\eta_0^3})~(\frac{\eta_0}{\eta})^{2(n-i)}~
\chi^{-2(n-i)+1} (\eta) +  \ldots = 0 \nonumber \\
&& \alpha, \beta = j\beta'_\Phi, \quad \quad j = -(n-1), 
\ldots, n-1 \label{chiselfevol}
\eea
where dots indicates subdominant terms. The effective potential in this equation includes negative 
power terms which can satisfy tracking condition if they vary slowly with time. 
Using the asymptotic expression of incomplete $\gamma$-function and counting the order of 
$\eta/\eta_0$ terms, we conclude that terms satisfying the following conditions vary slowly for 
$\eta/\eta_0 \gg 1$:
\be
\alpha = - 2\beta, \quad\quad 17 - 6n + 2i \geqslant 0 \label {powercount} 
\ee
The first condition eliminates the oscillatory terms, and the second one corresponds to orders of 
$\eta/\eta_0$ terms satisfying the tacking solution condition. As $i \leqslant n-1$, this condition 
is satisfied only for $n \leqslant 3$. The case of $n=4$ is also interesting, because although the 
indices of time-depending terms would be positive, the decay of the density of condensate would be 
enough slow such that its equation of state may be still consistent with observations. It is 
remarkable that these values for the self-interaction order are the only renormalizable polynomial 
potentials in 4-dimension spacetimes. The study of dark energy domination era is more complicated 
because the evolution equations of the condensate and expansion factor $a (\eta)$ become strongly 
coupled and must be solved numerically.

\subsection{Outline} 
In~\cite {p8,p49,p4}, we used non-equilibrium quantum field theory 
techniques to study the condensation of a scalar field during cosmological time. The scalar was 
assumed to be produced by the decay of a much heavier particle. Similar processes had necessarily 
happened during the reheating of the Universe. They could have happened at later times too if the 
remnants of the decay did not significantly perturb primordial nucleosynthesis. To fulfill this 
condition the probability of such processes had to be very small. We showed that one of the 
necessary conditions for the formation of a condensate is its light mass and small self-interaction 
which have important roles in the cosmological evolution of the condensate and its contribution to 
dark energy. In particular, we showed that only a self-interaction of order $\lesssim 4$ can 
produce a stable condensate in matter domination epoch. Confirmation of these results and the 
extension of the analysis to dark energy domination epoch needs lattice numerical calculation which 
is a project for near future. 

{\bf We conclude this section by reminding that if dark energy is the condensate of a scalar field, 
the importance of the quantum coherence in its formation and evolution would be the proof of the reign 
of Quantum Mechanics at largest observable scales of the Universe.} 

\section{Parametrization and test of dark energy models} \label{sec:deqft-param} % ~10 pages
\subsection {Introduction}
Modeling a physical phenomenon would not be useful if we cannot distinguish between candidate models. 
Specially, in what concerns the origin of accelerating expansion of the Universe, since its 
observational confirmation in the second half of 1990's, a large number of models are suggested to 
explain this phenomenon. In section \ref{sec:deqft-intro} we briefly reviewed the most popular 
categories of dark energy models. However, when it comes to their observational verification, the 
difficulty of the task oblige us to be more general, and at this stage only target the discrimination 
between three main category of dark energy: 
\begin{itemize}
\item Cosmological constant
\item Quintessence
\item Modified gravity
\end{itemize}
Fig. \ref{fig:dediscimcat} shows these categories and their possible impacts on various observables 
which can be potentially used to pin down or constrain the underlying model and discriminate it from 
other dark energy candidates. In fact, a notable difference between a cosmological constant, modified 
gravity and some of quintessence models is the presence of a weak interaction between matter and dark 
energy in the last two cases which can potentially leave, in addition to its effects on the large 
scale distribution of matter, other distinguishable imprints such as a hot/warm dark matter. 
Prospectives for multi-probe studies of dark energy is discussed in the next chapter.

\begin{wrapfigure}{r}{0.5 \textwidth}
\vspace{-0.5cm}
\includegraphics[width=9cm]{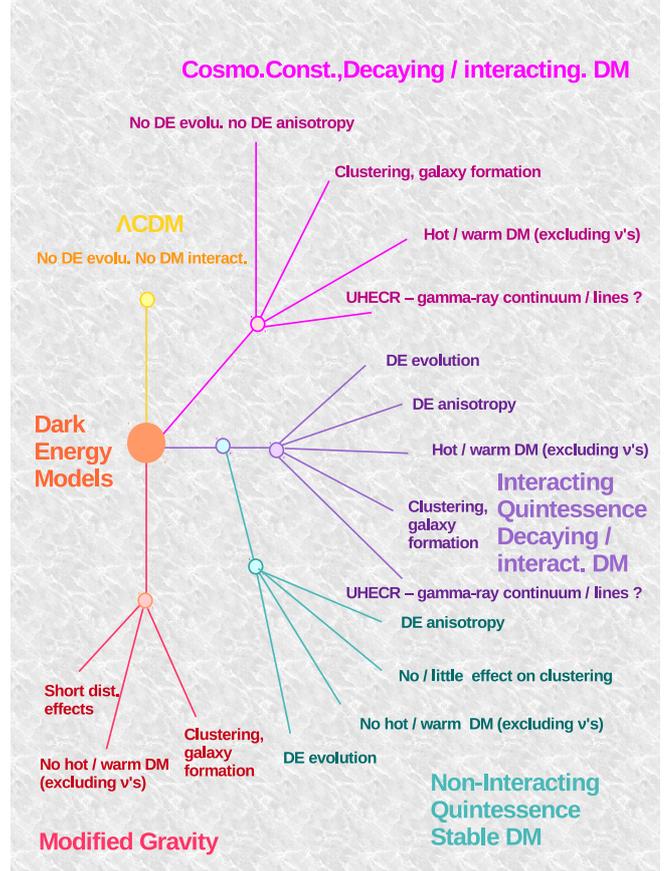}
\vspace{-1.5cm}
\caption{\sf \small The major categories of dark energy models and observables that potentially carry 
their imprint. A simpler version of this diagram is published in~\cite{p47}. Question marks 
mean that the effect is model dependent.\label{fig:dediscimcat}}
\vspace{-2cm}
\end{wrapfigure}

There are essentially two main cosmological observables which through their measurements cosmological 
parameters can be determined. The first observable is the expansion rate of the Universe - Hubble 
function $H (z)$ and its evolution with redshift. The second quantity is the distribution of matter 
anisotropies. The measurement of the first quantity needs a standard candle - an object with known 
luminosity or dimension. As for the second quantity, because most of matter in the Universe is dark, 
its distribution can only be measured indirectly through anisotropies that it induces in the 
distribution of Cosmic Microwave Background (CMB) and galaxies, or through its gravitational lensing 
effects. To be able to interpret measurements, specially for the purpose of discriminating among 
models, it is necessary to have quantitative descriptions for observables which can be universally 
applied to models irrespective of their details.

\subsection {Nonparametric determination of dark energy evolution}
Every content of the Universe has a contribution in the Friedmann equation that governs the evolution 
of expansion factor of the Universe: 
\be
H^2 (z) = (\frac{\dot{a}}{a})^2(z) = \frac{8\pi G}{3}\sum_i \rho_i \label{friedmanneq}
\ee
Therefore, the measurement of the expansion rate - the Hubble function - and its evolution are the most 
direct means for understanding homogeneous properties of dark energy. In a fluid approximation various 
contents of the Universe are characterized and distinguished by their equation of state $w$ defined in 
(\ref{wdefquin}). For a cosmological constant $w_{de} = -1$. This value can be considered as a critical 
point, because as we discussed in detail in section \ref{sec:deqft-int}, models with $w < -1$ apriori 
break the null energy theorem of general relativity, and therefore negative $w+1$ must be either an 
effective value or related to an exotic phenomenon such as a nonstandard kinetic term. For this 
reason, it is more useful to measure the sign of $w+1$, i.e. the direction of its deviation from the 
critical point, rather than its exact value which is less crucial for discriminating between models 
and more prone to measurement errors. The purpose of the work reported in~\cite{p15, p48,p47} was to 
find a suitable methodology to determine the sign of $w+1$ in a 
nonparametric manner. The expression {\it nonparametric} in signal processing literature means 
{\it testing a null hypothesis against an alternative by using a discrete condition such as a jump 
or the change of a sign rather than constraining a continuous parameter} (see e.g.~\cite{nonparam}). 
Therefore, for determining the sign of $w+1$ we need to find a quantity proportional to it 
irrespective of uncertainties of other parameters, as long as they are limited to a reasonable range.

In a flat universe containing cold matter, radiation and dark energy, all approximated by fluids, 
the density at redshift $z$ can be written as:
\be
\frac {\rho (z)}{\rho_0} = \Omega_m (1+z)^3 + \Omega_h (1+z)^4 + \Omega_{de} (1+z)^{3\gamma} 
\label{density}
\ee
where $\rho (z)$ and $\rho_0$ are respectively the total density at redshift $z$ and at $z=0$, 
$\Omega_m$, $\Omega_h$, and $\Omega_{de}$ are respectively the fraction of cold matter, radiation, and 
dark energy in the total density at $z=0$. For a constant $w$ (i.e. when it does not depend on $z$), 
$\gamma = w+1$, and it can be easily shown that in this case:
\be
{\mathcal A}(z) \equiv \frac{1}{3 (1+z)^2 \rho_0} \frac {d\rho}{dz} - \Omega_m = \gamma 
\Omega_{de}(1+z)^{3 (\gamma - 1)} \label {densderiv}
\ee
When various constituents of the Universe interact with each others, equations of states depend on $z$
and Friedmann equation and ${\mathcal A}(z)$ can be parametrized as the followings~\cite{p3}:
\be
\frac{H^2}{H_0^2} = \frac {\rho_c (z)}{\rho_{c0}} = \sum_i \Omega_i {\mathcal F}_{i}(z)
(1+z)^{3\gamma_i}, \quad \quad i=\text{{\it m, b, h,k,} and {\it de}} \label{friedmanintde}
\ee
where $m =$~cold dark matter, $b =$~baryons, $h =$~hot matter (radiation), $k =$~curvature, and 
$de =$~dark energy.
\be
\gamma (z) = \frac{1}{\ln (1+z)} \int_0^z dz'\frac{1 + w(z')}{1+z'} \label{gammade} 
\ee
\bea
{\mathcal B}(z) &\equiv& \frac{1}{3 (1+z)^2 \rho_0} \frac {d\rho}{dz} = \frac{-(2 \frac{dD_A}{dz} + 
(1+z) \frac{d^2D_A}{dz^2})}{\frac{2}{3(1+z)^2} (D_A + (1+z)\frac {dD_A}{dz})^3} \nonumber \\
&=& \sum_{i=m,h,k} \Omega_i \biggl (\gamma_i {\mathcal F}_i (z) + (1+z) \frac{d{\mathcal F}_i}{dz} 
\biggr )(1+z)^{3 (\gamma_i - 1)} + \Omega_{de}(w(z)+1)(1+z)^{3(\gamma_{de}(z) - 1)}\label{bzintdmde} \\
{\mathcal A}(z) &\equiv& {\mathcal B}(z) - \sum_{i=m,h,k} \Omega_i \gamma_i (1+z)^{3(\gamma_i - 1)} 
\nonumber \\
&=& \sum_{i=m,h,k} \Omega_i \biggl (\gamma_i ({\mathcal F}_i (z) - 1) + (1+z) 
\frac{d{\mathcal F}_i}{dz} \biggr ) (1+z)^{3 (\gamma_i - 1)} + \Omega_{de}(w(z)+1)
(1+z)^{3(\gamma_{de}(z) - 1)} \label{azintdmde}
\eea
It is clear that the sign of ${\mathcal A}(z)$ follows the sign of $\gamma$. Moreover, giving the 
fact that according to observations $\gamma \approx 0$, the exponent of $z$-dependent term in the 
r.h.s. of (\ref{azintdmde}) is always negative. This means that the maximum of ${\mathcal A}(z)$ is 
at $z \rightarrow 0$ where more precise data from standard candles such as supernovae type Ia are 
available. Another advantage of ${\mathcal A}(z)$ to direct determination of $\gamma$ from Friedmann 
equation is the fact that at low redshifts this equation is insensitive to the value of 
$\gamma$~\cite{p3}. In fact, using the definition of angular diameter distance $D_A (z)$, 
which in addition to supernovae data can be measured by Baryon Acoustic Oscillations (BAO), the 
Friedmann equation can be written as:
\be
\ln \biggl [\biggl (\frac{d}{dz}((1+z) D_A) \biggr)^{-1} - \Omega_m (1+z)^3 - \Omega_h 
(1+z)^4 - \Omega_K (1+z)^2\biggr ] = \ln \Omega_{de} + 3\gamma (z) \log (1+z) \label{loghubble}
\ee
At small redshifts the last term on the r.h.s. of (\ref{loghubble}) which contains $\gamma (z)$ 
approaches zero and its effect on the evolution of $D_A$ becomes negligibly small, irrespective of the 
value of $\gamma$. 

When $dw/dz \ll 3w(z)(w(z)+1)/(1+z)$, the sign of $dA/dz$ is opposite to the sign of $w(z)+1$. This 
condition is satisfied at low redshifts - see examples of models in Fig. \ref{fig:wzparam}. Therefore,  
$A(z)$ is a concave or convex function of redshift, respectively for positive or negative $w(z)+1$. 
Observations show that the contribution of $\Omega_k$ and $\Omega_h$ at low redshifts are much smaller 
than the uncertainty of $\Omega_m$. The function $dA/dz$ does not depend on $\Omega_m$. Thus, the 
uncertainty on the value of $\Omega_m$ can shift the value of $A(z)$ but it does not change its slope 
and its shape i.e. its concavity or convexity that determines the sign of $\gamma$ is preserved. 
Moreover, the uncertainty of $H_0$ scales ${\mathcal B}(z)$ uniformly at all redshifts and does not 
change geometrical properties of ${\mathcal A}(z)$. {\bf In conclusion, in what concerns the 
determination of the sign of $w+1$, the functions ${\mathcal A}(z)$ and ${\mathcal B}(z)$ are less 
sensitive to uncertainties of cosmological parameters than $H (z)$ and $D_A(z)$.}

Fig. \ref{fig:wzparam} shows ${\mathcal A}(z)$ for several phenomenological models of dark 
energy and parametrizations of $w(z)$. It is clear that for given values of ${\mathcal A}(z)$ 
(or similar quantities) at two redshifts the conclusion about evolution of dark energy and thereby 
the underlying model depends on the parametrization. Therefore, it is preferable to extract $w(z)$ in 
a nonparametric manner from data which is described in detail in~\cite{p3}.
\begin {figure}[h]
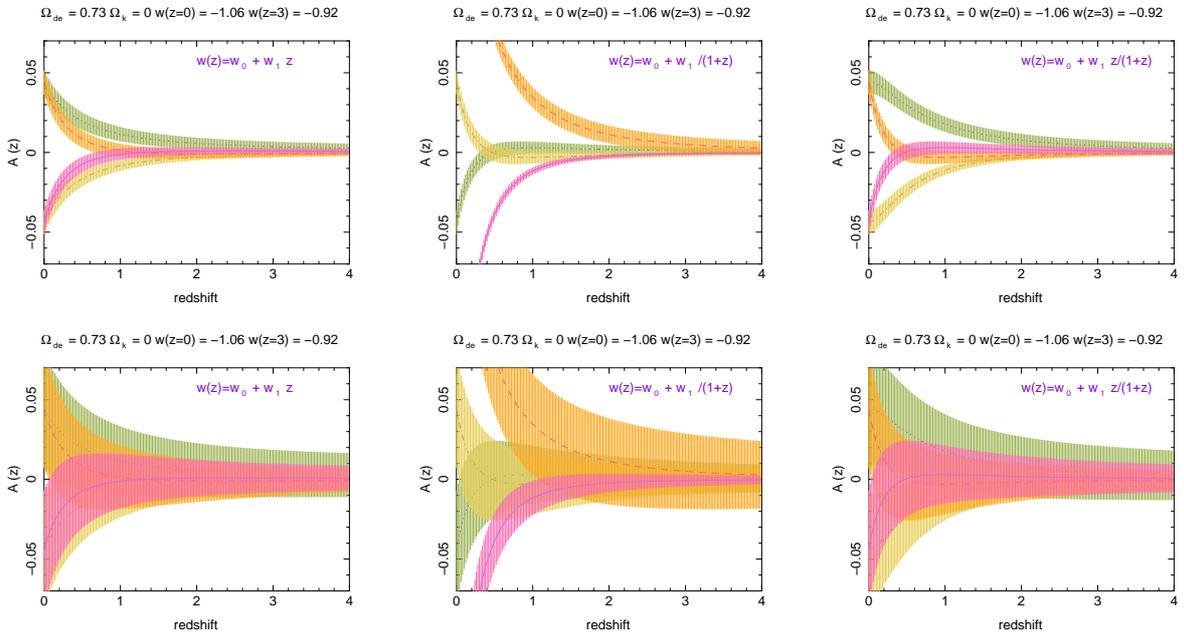

\begin {center}
\begin {tabular}{lll}
\includegraphics[width=4cm,angle=-90]{az1a.eps} &\includegraphics[width=4cm,angle=-90]{az1c.eps} & \includegraphics[width=4cm,angle=-90]{az1e.eps} \\
\includegraphics[width=4cm,angle=-90]{az1b.eps} &\includegraphics[width=4cm,angle=-90]{az1d.eps} & \includegraphics[width=4cm,angle=-90]{az1f.eps}
\end {tabular}
\end {center}
\caption{\sf \small ${\mathcal A}(z)$ as a function of redshift. To see how well ${\mathcal A}(z)$ can 
distinguish between various models and how systematic and statistical errors and parametrization 
affect the reconstructed model, we consider 3 parametrizations as written on the plots above. Note 
that parametrizations for the plot in the center and on the right are equivalent up to a redefinition 
of coefficients $w_0$ and $w_1$. All the examples have the same value of $w(z)$ at $z=0$ and $z=3$. 
After determining the corresponding coefficients $w_{0i}$ and $w_{1i}$ where index $i$ means initial. 
Then, to simulate systematic errors we have plotted the following models: $w_0 = -1 + |w_{0i} +1|$, 
$w_1 = w_{1i}$ (dot), $w_0 = -1-|w_{0i} +1|$, $w_1 = -w_{1i}$ (dot-dash), $w_0 = -1 + |w_{0i} +1|$, 
$w_1 = -w_{1i}$ (dash) and $w_0=w_{0i}$ and $w_1=w_{1i}$ (full). Colored vertical bars present 
statistical errors. The uncertainty of ${\mathcal A}(z)$ is $1\sigma_{{\mathcal A}(z=0)} = 0.01$ (top row) 
and $1\sigma_{{\mathcal A}(z=0)} = 0.05$ (bottom row) at $z=0$ and evolves with redshift as 
$\sigma_{\mathcal A}(z) = \sigma_{\mathcal A}(z= 0)(1+z)^2$. It seems to be possible to distinguish between 
normal and phantom dark energy models easily if uncertainties are limited to few percents. Evidently, 
achieving such a precision is challenging even for space missions such as Euclid. \label{fig:wzparam}}
\end {figure}

\subsubsection {Error estimation for nonparametric sign detection} 
Fig. \ref{fig:sncp} shows the comparison 
of ${\mathcal A}(z)$ calculated from supernovae data with dark energy models having positive and 
negative $w+1$. Visual inspection clearly concludes that both data-sets shown in this figure are 
consistent with $w+1 \lesssim 0$ in the interval $z \lesssim 0.5$. However, visual inspections or 
even the measurement of the slope lacks a quantitative estimation of uncertainties. In signal 
processing a binomial estimation of the probability or optimization of detection~\cite{signopt} are 
usually used to assess uncertainties. In fact, in most practically interesting contexts in signal 
processing the signal is constant and uncertainties are due to the noise. In the cosmological case 
discussed here the observable ${\mathcal A}(z)$ is both noisy and varies with redshift. Therefore, 
binomial probability and similar methods are not suitable. For this reason in~\cite{p15} another 
strategy which is specially appropriate for cosmological quantities is proposed. 

The null hypothesis for dark energy is $\gamma = 0$, i.e. $\Lambda$CDM. Assuming a Gaussian 
distribution for the uncertainties of the reconstructed ${\mathcal A}(z)$ from data and from simulated 
data with $\gamma = 0$, for each data-point we calculate the probability that it belongs to the null 
hypothesis. To include the uncertainty of data, we integrate the uncertainty distribution 1-sigma 
around its mean value:
\be
P_i = \frac{1}{\sqrt{2\pi (\sigma_{0i}^2 + \sigma_{i}^2)}} 
\int_{{\mathcal A}_i-\sigma_i}^{{\mathcal A}_i+\sigma_i} dx e^{-\frac {(x - {\mathcal A}_{0i})^2}
{2 (\sigma_{0i}^2 + \sigma_{i}^2)}} 
\label{probzero}
\ee
where ${\mathcal A}_i$ and $\sigma_i$ belong to the $i^{th}$ data-point, and ${\mathcal A}^{0i}$ and 
$\sigma_{0i}$ belong to the simulated null hypothesis model at the same redshift. Averaging over 
$P_i$ gives $\bar {P}$, an overall probability that the dataset corresponds to the null hypothesis. 
As $\gamma = 0$ is the limit case for $\gamma > 0$, $\bar {P}$ is also the maximum probability of 
$\gamma > 0$.
\begin{figure}[h]
\begin{center}
\vspace{0.5cm}
\begin{tabular}{cc}
\includegraphics[width=6cm]{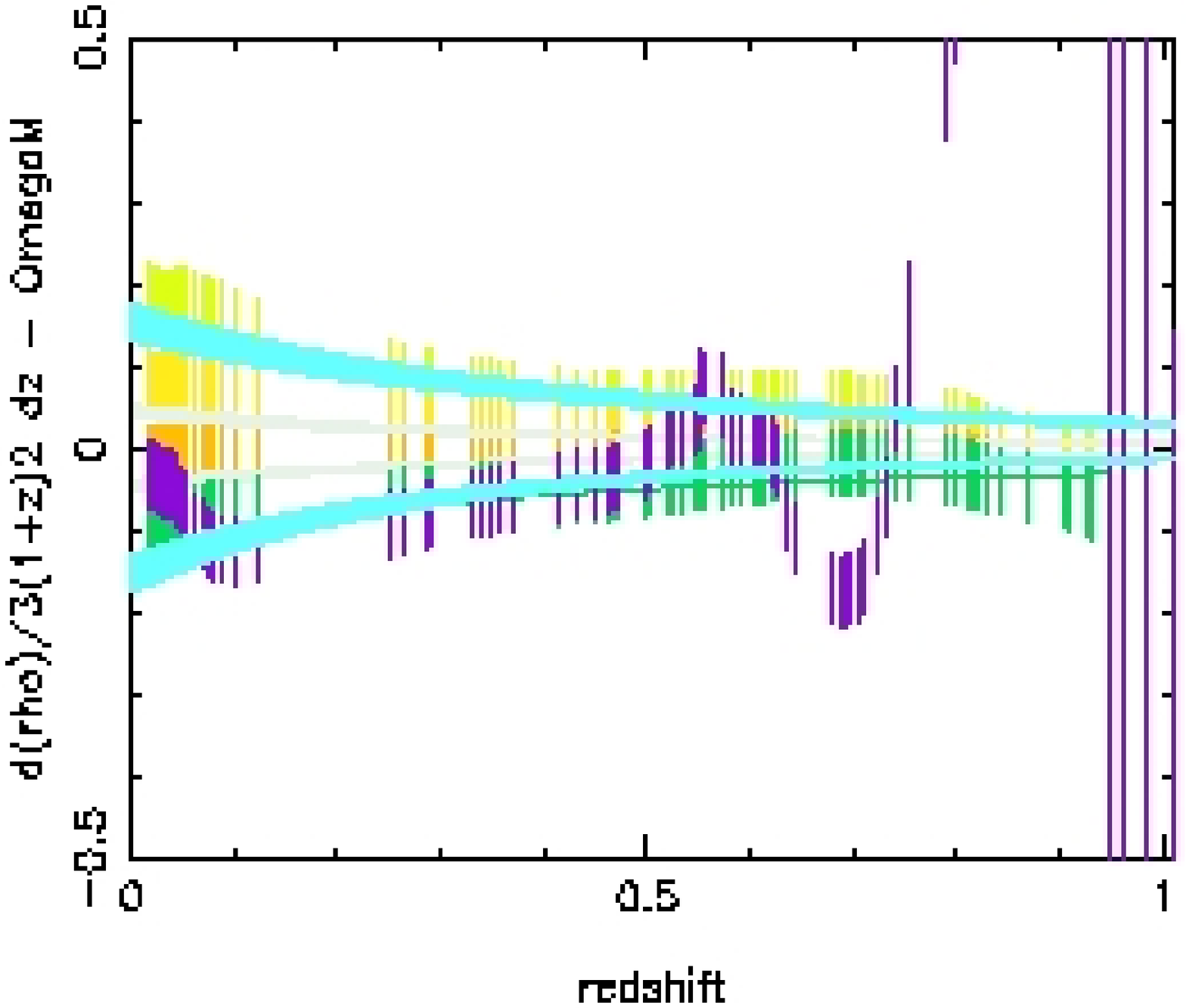} &
\includegraphics[width=6cm]{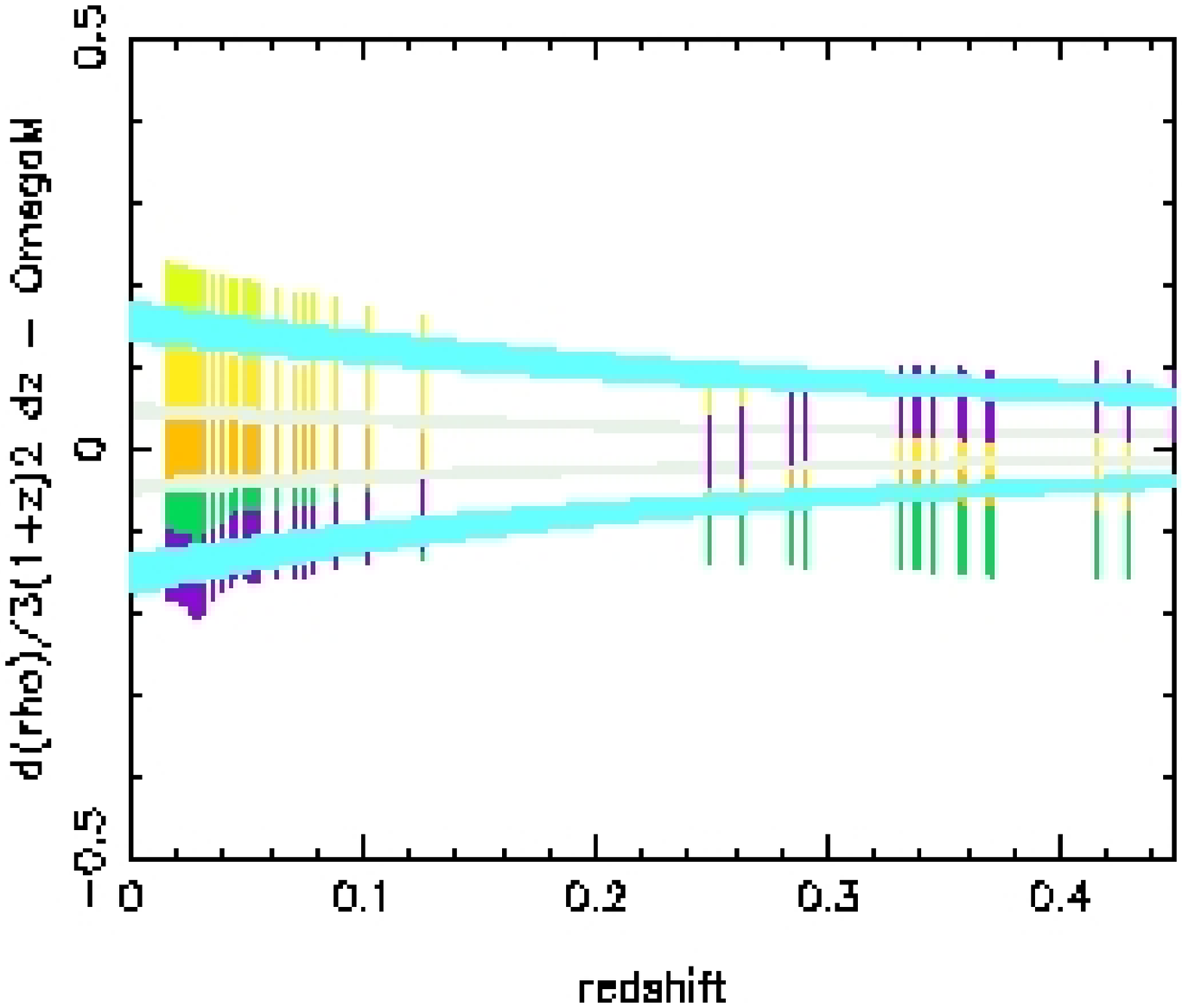}
\end{tabular}
\caption{\sf \small Left: ${\mathcal A}(z)$ from 117 supernovae from the SNLS survey (purple). 
Error bars are 1-sigma uncertainty. Green, orange, yellow, and light green curves are the 
reconstruction of ${\mathcal A}(z)$ from simulations for $\gamma = -0.2, -0.06, 0.6, 0.2$, 
respectively. The probability of null hypothesis ($\gamma = 0$) is $\bar{P} = 0.27$, therefore the 
probability of $\gamma < 0$, $1-\bar{P} = 0.73$. Light grey and cyan curves are theoretical 
calculation including the uncertainty on $\Omega_{de}$, respectively for $\gamma = \pm 0.06, \pm 0.2$. 
For all models $H_0 = 73~km~Mpc^{-1}~sec^{-1}$ and $\Omega_{de} = 0.77$ and $5\%$ errors for each. 
Right: ${\mathcal A}(z)$ for SNLS supernovae with $z < 0.45$. Definition of curves and cosmological 
parameters are the same as the left plot. For this dataset $1-\bar{P} = 0.93$. 
If $\Omega_{de} = 0.73$ is used, $1-\bar{P} = 0.96$. \label{fig:sncp}}
\end{center}
\end{figure}

\subsubsection {Precision of model discrimination}
Using generalized parametrization of Friedmann equation 
(\ref{friedmanintde}), in~\cite{p3} it is shown, with explicit examples, that discrimination 
between modified gravity and some categories of quintessence models is possible. Notably, if 
data is analyzed with the null hypothesis of no interaction, when in reality there is an interaction 
in the dark sector, the effective values of $\Omega_{de}$ and $\gamma_{de}$ extracted from measurements 
of the Hubble function $H(z)$ and from ${\mathcal A}$ will not be the same. Therefore, for detecting 
the tiny signature of an interaction in the dark sector, rather than fitting the data with a large 
number of parameters, it is better to measure the difference between two pair of measurements
$(\Omega_{eff}^{(H)},~\gamma_{eff}^{(H)})$ and $(\Omega_{eff}^{(A)},~\gamma_{eff}^{(A)})$. For this purpose, 
a natural criterion is:
\be
\Theta (z) \equiv \frac{\Omega_{eff}^{(A)} (w_{eff}^{(A)}(z) + 1) 
(1+z)^{3\gamma_{eff}^{(A)}(z)} - \Omega_{eff}^{(H)} (w_{eff}^{(H)}(z) + 1) 
(1+z)^{3\gamma_{eff}^{(H)}(z)}}{\Omega_{eff}^{(H)} (w_{eff}^{(H)}(z) + 1) 
(1+z)^{3\gamma_{eff}^{(H)}(z)}} \label{deftheta}
\ee
This quantity can be explained explicitly as a function of $\Omega_i,~{\mathcal F}_i,~$ and $\gamma_i$, 
for $i \in {m,~h,~k,~de}$. The variable $\Theta (z)$ is zero when $F_i= 1,~dF_i/dz = 0$. In addition, 
this expression 
can be used to determine the absolute sensitivity of a survey on an interaction in the dark sector, 
irrespective of the measured proxy and data analyzing methods. We should also remark that many 
authors have used quantities similar to ${\mathcal A}(z)$ that depend on the evolution of Hubble 
function $dH/dz$, see e.g.~\cite{simApara}. Nonetheless, the work presented in~\cite{p15,p48} is 
unique in proposing a nonparametric data analysing method for discriminating between dark energy 
models.

Assuming that $\Omega_m$ and $\Omega_h$ can be determined independently and with very good precision, 
for instance from CMB anisotropies with marginalization over $\gamma_{de}$, the quantity $\Theta$ may 
be determined from the measurement of $H(z)$ and ${\mathcal B}(z)$ by using data from whole sky or wide 
area spectroscopic surveys such as Euclid, or multi-band photometric surveys such as DES.

\subsection{Using LSS data for discriminating between dark energy models} \label{sec:lssdiscrim}
Fig. \ref{fig:dediscimcat} indicates that the deviation of matter clustering and spectrum of 
perturbations from predictions of $\Lambda$CDM are definitive signatures of interaction in the dark 
sector. Thus, along with the evolution of expansion rate discussed in the previous section, we must 
use LSS data to discriminate these models from $\Lambda$CDM and non-interacting quintessence. 
However, discrimination between modified gravity and interacting quintessence which both can be 
written as a scalar field model is not straightforward and more criteria are necessary. For this 
reason, in~\cite{p3} along with the studies reviewed in the previous section, we have 
investigated properties and differences of interacting quintessence and modified gravity to find 
new criteria for discrimination between them. Moreover, we suggest a new parametrization of 
observables for this purpose.

\subsubsection{Discrimination according to interaction type} By definition, in modified gravity it is 
expected that when it is written in Einstein frame, the scalar field has the same coupling to all 
species. However, as explained earlier this criterion is not suitable for observational and data 
analysing purposes, thus here we propose another criterion. But before explaining it, we review 
phenomenological description of interactions without a detail knowledge about the micro-physics of 
the underlying model.

In presence of non-gravitational interactions between constituents the energy-momentum tensor 
of each component $T^{\mu\nu}_i$ is not separately conserved, and conservation equations can be 
only written for the total energy-momentum tensor $T^{\mu\nu}$ defined as:
\bea
&& T^{\mu\nu} \equiv \sum_i T^{\mu\nu}_{i(free)} + T^{\mu\nu}_{int} \label{ttotdef} \\
&& T^{\mu\nu}_{;\nu} = \sum_i T^{\mu\nu}_{i(free)~;\nu} + T^{\mu\nu}_{int~;\nu} = 0 
\label{ttotcons}
\eea
where $T^{\mu\nu}_{i(free)}$ is the energy-momentum tensor of component $i$ in absence of interaction 
with other components. This means that $T^{\mu\nu}_{i(free);\nu} = 0$ only if the free value of fields 
and dynamical variables are used in this conservation equation. In this case 
$\sum_iT^{\mu\nu}_{i(int)~;\nu} = 0$. In perturbative field theories, specially when one studies the 
scattering of particles, it is assumed that in the out-of-interaction region particles are free and 
this formalism are applicable. But, in cosmology there is no {\it asymptotic freedom} because we 
live inside the interaction region. In the literature on interacting dark energy models 
(see e.g.~\cite{quinint,quinint0}) when only two constituents - matter and dark energy - are 
considered, the energy-momentum conservation equations are usually written as:
\be
T^{\mu\nu}_{m~;\nu} = Q^\mu, \quad T^{\mu\nu}_{\varphi~;\nu} = -Q^\mu \label{tqcons}
\ee
for an arbitrary interaction current $Q^\mu$. By comparing (\ref{ttotcons}) and with (\ref{tqcons}), 
it becomes clear that tensors in the left hand side of equations in (\ref{tqcons}) do not correspond 
to free energy-momentum tensors, and along with $Q^\mu$ they are obtained somehow arbitrarily by 
division of (\ref{ttotcons}). In fact, equations in (\ref{tqcons}) are inspired by perturbation 
theory in which for each perturbative order, the right hand sides of these equations are estimated 
using quantities from one perturbative order lower. Thus, they constitute an iterative set of 
equations from zero order (free) model in which $Q^\mu = 0$, up to higher orders. This approach is 
not suitable for dark energy where we ignore, not only interactions but also the free model. Therefore, 
in place of (\ref{tqcons}) we use the following general expression: 
\be
T^{\mu\nu}_{m~;\nu} = -Q_m^\mu, \quad T^{\mu\nu}_{\varphi~;\nu} = -Q_\varphi^\mu, \quad 
T^{\mu\nu}_{int~;\nu} = Q_m^\mu + Q_\varphi^\mu \label{tqmcons}
\ee
In these equations matter and dark energy tensors $T^{\mu\nu}_m$ and $T^{\mu\nu}_\varphi$ have the 
same expression as in the absence of interaction, but with respect to fields which are not free 
(dressed fields). These expressions can be justified by considering the effective Lagrangian. 
In Einstein frame the Lagrangian for a weakly interacting system can be divided to free and 
interaction parts:
\be
{\mathcal L} = \sum_i {\mathcal L}_i + {\mathcal L}_{int} \label{ltot}
\ee
Considering only local interactions, in the dynamics equations of fields the partial 
derivative of ${\mathcal L}_{int}$ with respect to each field determines the interaction term. 
Dynamics equations can be related to energy-momentum conservation equations 
(\ref{tqcons})~\cite{p25}. Therefore, interaction currents $Q_m^\mu$ and $Q_\varphi^\mu$ 
are generated by partial derivatives of ${\mathcal L}_{int}$ with respect to the corresponding 
interacting fields.

The scalar field in scalar-tensor modified gravity models is related to a dilaton. Consequently, the 
interaction term in these models is proportional to the trace of matter because it appears in the 
Lagrangian along with the metric. In this case, there is no interaction between the scalar field and 
relativistic particles, and it can be shown that $Q_m^\mu = -Q_\varphi^\mu$~\cite{frbean} and 
$T^{\mu\nu}_{int~;\nu} = 0$. The interaction current $Q^\mu$ for these models can be written as:
\be
Q^\mu = {\mathcal C}(\varphi) T_m \partial^\mu \varphi \label{mgcurrent}
\ee
where $T_m = g_{\mu\nu} T^{\mu\nu}_m$ is the trace of the energy-momentum tensor. In the literature 
interaction current of interacting dark energy models is usually considered to be like in 
(\ref{mgcurrent}). Therefore, we classify models with this type of interaction as modified gravity. 
In interacting quintessence models the interaction can be more diverse, notably it may depend on 
matter species. Below we present a phenomenological description for them without considering details 
of the underlying model. 

In quantum field theory, interactions can be easily included in the Lagrangian of the model. But 
this approach is usually useful if the microphysics of the model is studied. To be able to 
compare models with data, in observational cosmology we need macroscopic descriptions. For this 
reason, one usually uses fluid descriptions for fields. The transformation of the Lagrangian written 
with respect to fields to a fluid description is straightforward and energy-momentum tensor of 
interactions can be also described in fluid form without any ambiguity. However, their descriptions 
as a function of density and pressure of fluids depend on the details of interactions. For instance, 
a Higgs-like interaction between a scalar and a fermion $\propto \varphi \bar\psi \psi$ is described as 
$\propto (\rho_\psi - P_\psi)(\rho_\varphi - P_\varphi)^{1/2}$ if the self-coupling potential 
$V(\varphi) \propto \varphi^2$, and as $\propto (\rho_\psi - P_\psi) (\rho_\varphi - P_\varphi)^{1/4}$ if 
$V(\varphi) \propto \varphi^4$. Therefore, 
when the objective is a general parametrization of interactions without considering details of the 
underlying model, this type of description is not very suitable. 

A macroscopic description of the Lagrangian alone is not sufficient for describing microscopic 
processes which need the Boltzmann equation. In fact, it is well known that Boltzmann equation 
plays the role of the intermediate between quantum and classical description of interacting systems. 
Species are described by their phase space distribution $f(p,x)$ where $p$ and $x$ are respectively 
momentum and spacetime coordinates. Interactions are included in the equation as collisional terms, 
and one can obtain energy-momentum and number conservation equations directly by using properties of 
the Boltzmann operator, see e.g.~\cite{boltzcoll}. In the context of interacting dark energy models, 
the simplest examples of collisional terms are elastic scattering between dark matter and dark energy, 
and slowly decay of dark matter with a small branching ratio to dark energy similar to the model 
explained in section \ref{sec:deqft-int}\footnote{In models where energy is transferred from dark 
energy to dark matter, the interaction must be nonlinear and very sophisticated such that a very 
light quintessence field be able to produce massive dark matter particles. At present no fundamental 
description of interactions in these models is available.}. Finally, after some simplifying 
approximations, we find following expressions for the interaction terms of dark energy and dark matter 
component:
\bea
&& T^{\mu\nu}_{m~;\nu} \approx -\Gamma_m n_m^\mu + A_{ms} n_m^\mu u_{\varphi\rho} n_\varphi^\rho 
\equiv Q_m^\mu \label{tmconsmapprox} \\
&& T^{\mu\nu}_{\varphi~;\nu} \approx \Gamma_\varphi n_m^\mu + A_{\varphi s} n_\varphi^\mu 
u_{m\rho} n_m^\rho \equiv Q_\varphi^\mu \label{tdeconsmapprox}
\eea
where constants $\Gamma_i$ and $A_{is}$ are decay width and scattering amplitude for species $i$.
In the rest of this section we use these equations as an approximation for energy-momentum 
conservation equations irrespective of dark matter type (its spin) and details of interaction 
between two dark components. These details affect constants $\Gamma_i$ and $A_{is}$ which are used as 
parameters. One can also add dark matter self-annihilation term to (\ref{tmconsmapprox}). But, 
as it is proportional to $|n_m|^2$ its effect is significant only in dense regions and small spatial 
scales such as the central region of dark matter halos which are in nonlinear regime and were not 
studied in~\cite{p3}. 

\subsection {Parameters of background cosmology in interacting models} 
Now that we have interaction terms for the two main classes of interacting dark energy models we 
can determine ${\mathcal F}_i$ parameters defined in (\ref{friedmanintde}).
\begin{itemize}
\item {\bf Modified gravity} Using the energy momentum conservation equation (\ref{tqmcons}) and the 
interaction current for modified gravity models, the scalar field equation and the evolution equation 
of the homogeneous matter density are determined~\cite{frbean}:
\bea
&& \bar{\varphi}'' + 2{\mathcal H} \bar{\varphi}' + a^2 V_\varphi (\bar{\varphi}) = 
a^2 {\mathcal C} (\bar{\varphi}) \sum_i(\bar{\rho}_i -3 \bar{P}_i), \quad {\mathcal H} = 
\frac{a'}{a} \label {modgrphi}\\
&& \bar{\rho}'_i + 3{\mathcal H} (\bar{\rho}_i + \bar{P}_i) = {\mathcal C} (\bar{\varphi}) 
\bar{\varphi}' (\bar{\rho}_i - 3\bar{P}_i) \quad i=m,~b,~h\label {modgrrho}
\eea
where barred quantities are homogeneous components and $\varphi$ in subscript means derivative with 
respect to $\varphi$. Note that here we have generalized the original calculation in~\cite{frbean} by 
considering a $\varphi$-dependent ${\mathcal C} (\bar{\varphi})$ coefficient in the right hand side 
of these equations rather than the constant value of models ${\mathcal C} = \sqrt {4\pi G/3}$ for 
$\mathrm {f}(R)$ models~\cite{frbean}. Equations (\ref{modgrphi}) and (\ref{modgrrho}) are coupled 
and an analytical solution cannot be found without considering an explicitly $V(\varphi)$. 
Therefore, to solve the equation for $\bar{\rho}$, which is in fact the only directly 
observable quantity, we simply consider the right hand side of the equation as a time-dependent 
source. The solution of equation (\ref{modgrrho}) can be written as:
\be
\bar{\rho}_i (z) = \bar{\rho}_i (z_0) (1+z)^{3(1+w_i)} e^{(1-3w_i) F(\bar{\varphi})}, \quad 
F(\varphi) \equiv \int {\mathcal C} (\bar{\varphi}) d\varphi, \quad i = m,b,h \label {mgrhohomo}
\ee 
where $w_i \equiv \bar{P}_i/\bar{\rho}_i$ for all species except dark energy are assumed to be 
constant. Comparing this solution with (\ref{friedmanintde}) we find:
\be
{\mathcal F}_i(z) = e^{(1-3w_i) F(\bar{\varphi} (z))} \approx 1 + (1-3w_i) F(\bar{\varphi} (z)) 
\label{mgfi}
\ee
\item {\bf Interacting quintessence} In the same way, we can determine ${\mathcal F}_i$ coefficients 
for (interacting)-quintessence using equation (\ref{tmconsmapprox}). After taking some 
approximations discussed in detail in~\cite{p3}, the evolution equation of densities in  
interacting quintessence models becomes:
\be
\bar{\rho}'_i + 3{\mathcal H} (\bar{\rho}_i + \bar{P}_i) = -\Gamma_i a \bar{\rho}_i + A_{si} a 
\bar{\rho}_i \bar{\rho}_\varphi \label{quinrhohomoeq}
\ee
where $i$ indicates any cold matter or relativistic species that interact with quintessence field. 
A clear difference between interaction term in (\ref{quinrhohomoeq}) and (\ref{modgrrho}) is that the 
former does not explicitly depend on the scalar field, and therefore we do not need to know and solve 
a field equation similar to (\ref{modgrphi}). The solution of this equation and corresponding 
${\mathcal F}_i$'s are:
\bea
\bar{\rho}_i (z) &=& \bar{\rho}_i (z_0) (1+z)^{3(1+w_i)} \exp \biggl (\Gamma_i (\tau (z) - 
\tau (z0)) + A_{si}\int dz \frac{\bar{\rho}_\varphi (z)}{(1+z) H (z)} \biggr ) 
\label{quinrhohomo} \\
{\mathcal F}_i (z) &=& \exp \biggl (-\Gamma_i (\tau (z) - \tau (z_0)) + A_{si}\int dz 
\frac{\bar{\rho}_\varphi (z)}{(1+z) H (z)} \biggr ) \approx 1 + \Gamma_i (\tau (z_0) - \tau (z)) + 
\nonumber \\
&& A_{si}\int_{z_0}^z dz \frac{\bar{\rho}_\varphi (z)}{(1+z) H (z)} \label{quinfi}
\eea
where $\tau (z)$ is the age of the Universe at redshift $z$. Note that even in absence of expansion, 
the density of dark matter at high redshifts (large distances from us) is higher if $\Gamma_m > 0$.
\end{itemize}
Along with consistency relation explained above for modified gravity models, explicit dependence 
of (\ref{quinfi}) on measurable quantities $\bar{\rho}_\varphi (z)$ and $H (z)$ apriori allows to 
discriminate between interacting quintessence and modified gravity models. However, the prior 
knowledge about the evolution of these quantities are mandatory for distinguishing the underlying 
model and without such information one cannot single out any model.

\subsection {New parametrization of perturbations in interacting dark energy models} 
The criteria proposed in this section for distinguishing between modified gravity and interacting 
quintessence models are based on their interactions. Therefore, we expect different evolution for 
matter anisotropies and dark energy density in these models. In fact, if we could decompose the 
interaction current to terms proportional to scalar metric perturbations and matter density 
fluctuations, it were possible to distinguish between models easily. However, in practice measured 
quantities are the power spectrum of matter perturbations and its growth rate $\mathbf{f}(z,k)$ defined 
as:
\be
\mathbf{f}(z,k) \equiv \frac{d\ln D}{d\ln a} = \frac{\delta'_m}{{\mathcal H} \delta_m}, \quad 
D \equiv \frac{\delta_m (z,k)}{\delta_m (z=0,k)} \label{growthrate}
\ee
The function $\mathbf{f}(z,k)$ is usually extracted from the power spectrum using a 
model~\cite{fextect0}, for instance a power-law for the primordial spectrum, modified to include 
Kaiser effect and redshift distortions due to the velocity dispersion. 

To obtain the evolution equation of $\mathbf{f}(z,k)$, we replace gravitational potentials $\psi$ and 
$\phi$ by expressions depending only on $\delta_m \equiv \delta \rho_m / \bar{\rho}_m$ and 
$\theta_m \equiv ik_jv_{(m)}^j$. Assuming a negligible anisotropic shear at 
$z \lesssim \mathcal {O} (1)$ which concerns galaxy surveys, scalar metric perturbations $\psi$ and 
$\phi$ can be determined from Einstein equations:\footnote{In this section for the sake of 
simplicity of notation we consider that ${\mathcal F}_i$'s factors for species are included in 
$w_i$'s, i.e. $(1+z)^{3\gamma_i} {\mathcal F}_i$ is redefined as $(1+z)^{3\gamma_i(z)}$ and $w_i$ 
is obtained from (\ref{gammade}) using this redefined $\gamma_i$. Therefore, for interacting dark 
energy models $w_m$ is nonzero and in general depends on redshift.}
\bea
ds^2 & = & a^2(\eta) \biggl [(1 + 2\psi) d\eta^2 - (1 - 2\phi) \delta_{ij} dx^i dx^j \biggr ] 
\label{metricaniso} \\
\phi = \psi &=& \frac{4\pi G \bar{\rho}_m}{k^2} \bigg (\delta_m + 3 (1+w_m) \frac{{\mathcal H}
\theta_m}{k^2} \biggr ) + \Delta \psi \label{phipsisol} \\
%\Delta \psi &=&  \frac{4\pi G}{k^2} \biggl (\delta\rho_\varphi - 3 {\mathcal H} \delta \varphi (\bar{\rho}_\varphi + \bar{P}_\varphi)^{\frac{1}{2}}\biggr ) \label{deltapsi} \\
\phi' &=& -\frac{4\pi G \bar{\rho}_m {\mathcal H}}{k^2} \biggl (\delta_m + 
(3 + \frac{k^2}{{\mathcal H}^2}) (1+w_m) \frac{{\mathcal H} \theta_m}{k^2} \biggr ) + 
\Delta \phi' \label{phip} \\
%\Delta \phi' &=& -{\mathcal H}\Delta \psi + 4\pi G a^2 \delta \varphi (\bar{\rho}_\varphi + \bar{P}_\varphi)^{\frac{1}{2}}  \label{deltaphip}
\eea
In (\ref{phipsisol}) and (\ref{phip}) terms that vanish for $\Lambda$CDM model are included in 
$\Delta \psi$ and $\Delta \phi'$. They can be described as a function of two new quantities 
$\epsilon_0$ and $\epsilon_1$:
\bea
&& \epsilon_0 \equiv \frac{\delta \rho_\varphi}{\bar{\rho}_m}, \quad \epsilon_1 \equiv 
\frac{{\mathcal H} (\bar{\rho}_\varphi + \bar{P}_\varphi)^{\frac{1}{2}} \delta \varphi}
{\bar{\rho}_m} \label{epsilo01def} \\
&& \Delta \psi = \frac{4\pi G \bar{\rho}_m}{k^2} (\epsilon_0 -3 \epsilon_1), \quad \quad  
\Delta \phi' = -\frac{4\pi G \bar{\rho}_m {\mathcal H}}{k^2} \biggl (\epsilon_0 - 
(3 + \frac{k^2}{{\mathcal H}^2}) \epsilon_1 \biggr ) \label{deltaphipredef}
\eea
After replacing $\phi'$ and $\psi$ in the evolution equation of matter and velocity perturbations 
we can determine the evolution of growth rate which can be directly measured from galaxy distribution:
\bea
&& \mathbf{f}' {\mathcal H} + \mathbf{f} ({\mathcal H}' + {\mathcal H}^2) + 
\mathbf{f}^2 {\mathcal H}^2 + 3 (C_{sm}^2 - w_m) ({\mathcal H}' + \mathbf{f} {\mathcal H}^2) + 
3 (C_{sm}^2 - w_m) {\mathcal H}^2 + \nonumber\\
&& \quad \quad \quad \frac{3}{2}~\Omega_m (1+w_m)^2 {\mathcal H}^2 + k^2 C_{sm}^2 + 
E_0 \mathbf{f} {\mathcal H} + E_1 k^2 + E_2 {\mathcal H} + E_3 {\mathcal H}^2 + E_4 = 0 
\label{growthratevol}
\eea
Coefficients $E_0,~E_1,~E_2,~E_3,~E_4$ depend on $z$, $k$, equation of state of matter, sound speed, 
and parameters related to the interaction in the dark sector. The details of their expression can 
be found in~\cite{p3}. For $\Lambda$CDM model $E_i = 0,~i=0,\cdots,4$. For a non-interacting 
quintessence model all $E_i$ coefficients are zero except $E_3$. A notable difference between 
modified gravity and interacting quintessence models is the coefficient $E_1$ which is strictly 
zero for interacting dark energy models and nonzero for modified gravity which leaves an additional 
scale dependent signature on the evolution of matter anisotropies. The other explicitly scale 
dependent term is common among all models and is expected to be very small because it is proportional 
to the square of sound speed which is very small for cold matter. In addition, in contrast to other 
$E_i$ coefficients, $E_1$ and $E_3$ are dimensionless. Evidently, the contribution of $E_1 k^2$ term 
in equation (\ref{growthratevol}) with respect to other terms increases with $k$, i.e. at shorter 
distances. But, nonlinearity effects such as mode coupling also increase at large $k$. They can 
imitate an interaction in the dark sector and lead to misinterpretation of data. For this reason, 
simulations show that observations of galaxy clusters is a good discriminator between dark energy 
models~\cite{mgcluster}, because they are still close to linear regime, but have relatively large $k$.

Discriminating quality of a survey can be estimated by the precision of $E_1$ and $E_3$ measurements. 
However, one expects some degeneracies when equation (\ref{growthratevol}) is fitted 
to determine $E_i$'s. Moreover, in galaxy surveys, $\mathbf {f}$ and $\mathbf {f}'$ (or more 
precisely $d\mathbf {f} / dz$) are determined from the measurement of the power spectrum, itself 
determined from the galaxy distribution, and ${\mathcal H}$ and ${\mathcal H}'$ are determined from 
the BAO feature of the spectrum. Thus, these measurements are not completely independent. An 
independent measurement of ${\mathcal H}$ and ${\mathcal H}'$ e.g. using supernovae will help to reduce 
degeneracies and error propagation from measured quantities to the estimation of $E_i$'s. The relation 
between ${\mathcal H}'$ and $B(z)$ shows the logical connection between the parametrization of 
background cosmology and the evolution of fluctuations, specially in what concerns discrimination 
between dark energy models. In fact, anisotropies depend on the equation of state of matter, 
which in the framework of interacting dark energy models, is modified by its interaction with dark 
energy. Thus, their independent measurements optimize their employment in the procedure of 
distinguishing among various models.

\subsection {Interpretation and comparison with other parametrizations}
Definition of parameters $\epsilon_0$ and $\epsilon_1$ in (\ref{epsilo01def}) show that the 
former depends only on dark energy density anisotropies and the latter only on the peculiar velocity 
of dark energy field, i.e. on its kinematics. They follow each other closely and approach zero when 
the field approaches its minimum value. However, their exponent close to the minimum depends on the 
interaction. Therefore, their measurements give us information about the potential and interactions 
of the scalar field. Moreover, the difference in the dependence of the evolution equation of 
anisotropies and growth factor to these parameters shows that only by separation of kinematics 
and dynamics of dark energy it would be possible to distinguish between modified gravity and other 
scalar field models.

The deviation of gravity potentials $\phi$ and $\psi$ from their value in $\Lambda$CDM 
$\Delta\psi$ is the quantity which can be measured directly from gravitational lensing data. For 
this reason, various authors have used $\Delta\psi$ to parametrize the deviation of models and data 
from $\Lambda$CDM~\cite{param0}. However, equations (\ref{phipsisol}) and (\ref{deltaphipredef}) show 
that although $\Delta\psi \neq 0$ is by definition a signature of deviation from $\Lambda$CDM, 
in contrast to claims in the literature, it is not necessarily the signature of a modified gravity 
because quintessence models, both interacting and non-interacting, also induce $\Delta\psi \neq 0$. 

Because we have used Einstein frame for both quintessence and modified gravity models, in 
absence of an anisotropic shear $\phi = \psi$ even in models other than $\Lambda$CDM. At linear order, 
gravitational lensing effect depends on the total potential $\Phi \equiv \phi + \psi$. Therefore, 
in Einstein frame:
\be
\Phi = 2\phi = 2\psi = \Phi_{\Lambda\text{CDM}} + 2\Delta\psi, \quad \quad 
\Phi_{\Lambda\text{CDM}} \equiv \frac{4\pi G \bar{\rho}_m}{k^2} \bigg (\delta_m + 3 (1+w_m) 
\frac{{\mathcal H}\theta_m}{k^2} \biggr ) \equiv \frac{4\pi G \bar{\rho}_m}{k^2 \Delta_m} 
\label{totpotent}
\ee
In the notation of~\cite{param0} $\Phi = 2\Sigma \Phi_{\Lambda\text{CDM}}$, thus: 
\be
\Sigma = 1 + \frac{\Delta\psi}{\Phi_{\Lambda\text{CDM}}} = \frac{\epsilon_0 - 3\epsilon_1}
{k^2 \Delta_m} \label{sigmalens}
\ee
The other quantity which affects the evolution of lensing and directly depends on the cosmology is 
the growth factor of matter anisotropies which determines the evolution of $\Delta_m$ defined 
in (\ref{totpotent}). This quantity can be obtained from integration of growth rate $\mathbf{f}$ 
defined in (\ref{growthrate}) and is usually parametrized as $\Omega_m^{\gamma}$. For $\Lambda$CDM 
$\gamma \approx 0.55$~\cite{growthparam}. In this respect there is no difference 
between our formulation and what is used in the literature. 

Our parametrization can be related to parameters $\eta$ and $Q$ used in the 
literature~\cite{param0}: $\eta \equiv (\psi - \phi)/\phi$ and $Q = \phi/\phi_{\Lambda\text{CDM}}$. Thus, 
parameters $\Sigma$ and $\eta$ are not independent and $\Sigma = Q (1+\eta/2)$. In Einstein frame 
$\eta = 0$ unless there is an anisotropic shear. At first sight it seems that there is less information 
in Einstein frame about modified gravity than in Jordan frame. However, one should notice that 
in Einstein frame the fundamental parameters are $\epsilon_0$ and $\epsilon_1$ and other 
quantities such as $\Delta\psi$ and $\mathbf{f}$ can be expressed with respect to them. Therefore, 
the amount of information in Einstein and Jordan frames about modified gravity is the same. 
The advantage of the formulation in Einstein frame and parameters $\epsilon_0$ and $\epsilon_1$ is 
that they can be used for both quintessence and modified gravity. Moreover, they have explicit 
physical interpretations that can be easily related to the underlying model of dark energy. 
Comparison of our parametrization with few others proposed recently can be found in~\cite{p3}. 
In our knowledge no other parametrization of the gross factor comparable to $E_i$'s exist in the 
literature.

\subsection{Outline}
We proposed a nonparametric formalism to determine the sign of $w+1$ in the equation of state of 
dark energy which has crucial importance for discrimination between a cosmological constant, 
quintessence and phantom dark energy models. We showed that it is a better discriminator and 
less sensitive to uncertainties of other cosmological parameters and the noise in the data than 
fitting procedures of continuous parameters.

We parametrized the evolution of homogeneous and perturbations of the constituents of the Universe 
for modified gravity and interacting quintessence models. We have showed that when the interaction 
is ignored in the data analysis, the effective value of parameters are not the same if we calculate 
them from Friedmann equation or from a function proportional to the redshift evolution of the total 
density. We defined a quantity that evaluates the strength of the interaction. Its observational 
uncertainty can be used to estimate the quality of cosmological surveys in what concerns their ability 
to discriminate among dark energy models. Additionally, we obtained a new parametrized description 
for the evolution equation of the growth rate of matter anisotropies which can be used for 
discriminating between $\Lambda$CDM, modified gravity and interacting quintessence models. 

\section{A note on the backreaction of perturbations as the origin of dark energy} \label{sec:backreact}
In Fig. \ref{fig:demodel} there is a separate set of models called {\it dark energy as a 
back-reaction}. Their common aspect is that they associate the accelerating expansion of the Universe 
not to a new energy content but to the back-reaction of anisotropies of matter distribution assumed 
not being properly taken into account in cosmological models~\cite{brkolb,brkolb0,brbuchert}. Many 
authors have commented against these models, see e.g.~\cite{responpap}. \cite{p45} gives a 
non-technical summary of main arguments against the possibility of an enough large back-reaction 
being able to explain the observed dominant contribution of dark energy in the expansion of the 
Universe.

Our argument against the claims of~\cite{brkolb,brkolb0} is very simple. Very small amplitude of 
the CMB anisotropies which include the largest accessible scales to electromagnetic probes, proves 
that at large scales the deviation from homogeneity of matter and energy distribution is very small. 
In this case to determine the average expansion rate of the Universe we can expand the two sides of 
the Einstein equations with respect to fluctuation scale (or the spectrum in Fourier space):
\be
\langle G_{\mu\nu}^{(0)}\rangle = 8\pi G \langle T^{(0)}_{\mu\nu} \rangle + 
8\pi G \langle T^{(1)}_{\mu\nu} + T^{(2)}_{\mu\nu} + \ldots \rangle - 
\langle G^{(1)}_{\mu\nu} + G^{(2)}_{\mu\nu} + \ldots \rangle
\ee
It is crucial to emphasis that in this expansion by definition:
\be
\mathbf {\langle T_{\mu\nu} \rangle = T_{\mu\nu}^{(0)}} \quad \quad \text{or} \quad \quad 
\mathbf {\langle G_{\mu\nu} \rangle = G_{\mu\nu}^{(0)}} \label{zerordereinstein}
\ee
The validity of one of the conditions in (\ref{zerordereinstein}) is necessary because we must have 
a reference point with respect to which we determine deviations from homogeneity. We have presented  
both relations because their application depends on the physical quantities which are measured. 
The first condition is applied when the distribution of matter is observed. The second one must be 
applied when the expansion rate i.e. the geometry of the Universe is measured. Under the assumption 
of perturbative fluctuations:
\bea
&& {\mathcal K}_{\mu\nu} \equiv 8\pi G \langle T^{(1)}_{\mu\nu} + T^{(2)}_{\mu\nu} + 
\ldots \rangle \sim {\mathcal O}(\epsilon) \rightarrow 0, \quad \quad {\mathcal K}_\mu^\mu \sim 
\delta \equiv \frac{\delta \rho}{\rho} \ll 1 \label{matterfluct} \\
&& \langle G^{(0)}_{\mu\nu} + 
G^{(1)}_{\mu\nu} + \ldots \rangle = 8\pi G \langle T^{(0)}_{\mu\nu} + T^{(1)}_{\mu\nu} + 
\ldots \rangle \label{einsteinfluct}
\eea
The expression (\ref{matterfluct}) is the necessary condition for a perturbative expansion, and its 
applicability here is in accord with observations. Based on (\ref{zerordereinstein}) and mathematical 
definition of a perturbative expansion, the terms of the same order in the two sides of the equation 
(\ref{einsteinfluct}) must be equal. In this formulation the error made by considering zero-order 
terms in (\ref{einsteinfluct}) is at most of order $\delta \ll 1$ which cannot explain a dark energy 
roughly 2.5 folds larger than matter component.

However, in equation (10) of~\cite{brkolb} the condition (\ref{zerordereinstein}) is violated:
\be
G^{(0)}_{\mu\nu} = 8\pi G \langle T_{\mu\nu} \rangle - \langle G^{(1)}_{\mu\nu} \rangle. 
\label{kolbperturb}
\ee
In this expansion it is not clear how perturbative terms are related to nonperturbative quantities. 
Evidently, in this case it would be always possible to choose {\it perturbative terms} such that 
the term $\langle G^{(1)}_{\mu\nu} \rangle$ be enough large to explain the acceleration of the Universe. 
The caveat in the above argument can be the effect of limited volume of observations, at most as 
large as the horizon. Based on this issue, it was claimed that super-horizon modes can explain the 
apparent accelerating expansion. Moreover, because at small scales $\delta \gg 1$, other 
authors~\cite{brbuchert} have claimed that accelerating expansion can be due to error in averaging 
sub-horizon anisotropies. 

The average value of an arbitrary scalar quantity $\psi (X,t)$ in a constant-time volume $V_D$ is defined 
as~\cite{brbuchert}:
\be
\langle \psi \rangle_D \equiv \frac{1}{V_D} \int_{V_D} d^3 X J \psi, 
\quad \quad J = \sqrt{|\det g_{ij}|}, \quad \quad V_D \equiv 
\int_{V_D} d^3 X J \label{aver}
\ee
Therefore, a finite volume has the effect of a spherical window function, and the average value of 
anisotropies can be expressed as:
\bea
&& \langle \delta^2 \rangle_D = \int_D d^3 x J \delta^2 (x) = \int d^3 k P (k) sinc (k_i x^i_D) 
\label{eqtimeavaniso}
\\
&& \Delta \equiv \langle \delta^2 \rangle_D - \langle \delta^2 \rangle_\infty = 
\int d^3 k P (k) sinc (k_i x^i_D) - P (k=0) \label{eqtimestdaniso}
\eea
where $x_D^i$ is a characteristic size scale of the volume $V_D$ in the direction $x^i$. For instance, 
for a cube parallel to the coordinate axes, it is the length of the edge parallel to axis $i$. 
When the spectrum of inhomogeneities is scale-independent, from properties of $sinc$ function we 
can conclude that the contribution of modes $k \gg 1/x_D$ i.e. inhomogeneities at scales much smaller 
than $X_D$ are negligible. In fact in~\cite{brbuchert} it is proved that only in a highly curved 
universe these modes can induce a large constant term similar to dark energy. 

The contribution of each mode in the range of $0 < k < 1/X_D$ is proportional to $1/k$. Therefore, 
in the case of a scale-independent spectrum where statistically averaged value of 
$\delta (k)$ is mode independent, the integral over these modes after renormalization of IR 
divergence is a sub-dominant logarithmic term $\propto \log x_D^{-1}$ and in an inflationary universe 
where $x_D \rightarrow \infty$ is very small. This confirms the results of~\cite{responpap} and shows 
that the right hand side of:
\be
 \langle \partial_t \delta \rangle_D - \partial_t \langle \delta \rangle_D = 
\langle \theta \delta \rangle_D - \langle \theta \rangle_D \langle \delta \rangle_D \label {timevar}
\ee
which has been claimed to be the origin of apparent accelerating expansion, is very close to zero. 
Therefore, the assumption of commutation between time and space averaging is a good approximation. 
Indeed, the difference between integration in a finite and an infinite volume decreases with the 
expansion of the Universe, in contrast to dark energy behaviour which becomes more dominant with time. 
Note that the argument given here is only based on the statistical properties of inhomogeneities and 
perturbative anisotropies has not been assumed. Many authors have gone to lengthy demonstration to 
show that the effect of backreaction is small. Nonetheless, we believe that (\ref{kolbperturb}) is the 
origin of confusion and falsifies all the calculations based on this ambiguous expansion. This point 
was also remarked and raised by other authors see e.g.~\cite{quinpertbackreact}.

\section{Conclusion} \label{sec:deconclude}
In this review we presented various aspects of the origin of the accelerating expansion of the 
Universe and fundamental problems that it imposes on our understanding of physics and cosmology. 
By considering the problem both from fundamental and quantum field theoretical view, and classically, 
we showed that some of the suggested explanations such as vacuum energy or backreaction of 
anisotropies have serious difficulties to be the cause of the observed phenomenon. We reviewed 
a new vacuum state in which the expectation value of number operator for all modes is zero 
independent of the spacetime reference frame used for definition of modes. The presence of such state 
assures that it makes sense to talk about an {\it empty} space in quantum field theory, although it 
may be an abstract state that never existed. On the other hand, the Universe may be filled by 
condensates of quantum fields which symmetries prevent to acquire a large effective mass, and thereby 
they have quintessence-like behaviour. Phenomenological and non-equilibrium quantum field theoretical 
study of these models show that they can replace a cosmological constant and may have late time 
behaviour very close to the latter. Only additional insight to particle physics of the dark sector may 
provide sufficient criteria to distinguish between these models and a simple cosmological constant. 

A subject which is not covered in this review is the adiabatic instability of some interacting 
quintessence models~\cite{quininstability} which constrains some interacting dark energy models, 
notably quintessence models with negative exponential potential. We have already 
discussed that models with such a potential are not physically motivated, because despite having small 
couplings, they do not have a scattering interpretation. On the other hand, the class of models 
discussed in this review can have polynomial potentials of positive order with well defined scatting 
interpretation. Moreover, the relation between density of dark matter and amplitude dark energy and 
their effect on the expansion rate induces a self-control process which suppresses instabilities.

Further investigations, both theoretical and numerical simulations, are necessary the understand 
quantum physics of quintessence models, and its relation with inflation. On the theoretical side, 
it is necessary to extend the formulation of the evolution of condensate to full Kadanoff-Baym 
equations - 2-Particle Irreducible (2PI) expansion, and should include the evolution of expansion rate 
in a consistent manner. Moreover, calculations must be extended to the epoch of dark energy dominance 
at redshifts $\lesssim 1$ which corresponds to about half of the age of the Universe. Such a study is 
crucial for understanding the nature of dark energy because during this epoch the expansion of the 
Universe becomes even faster and more challenging for the survival of a quintessence condensate. The 
study of such a complicated formulation is only possible through numerical simulations. analytical 
results reviewed in this article could be obtained only under various simplifying approximations. 
This project is a work in progress~\cite{houricondqm}. Finally, more realistic 
models rather than simple toy models reviewed here should be investigated and detectable accelerator 
and astro-particle physics signatures should be identified.

Models considered here for studying the evolution of condensates are prototypes of processes that, 
according to our present understanding of the physics of the early Universe, had occurred irrespective 
of the unknown details of high energy particle physics. One of the most notable outcome of these 
studies is the significant role of quantum nature of dark energy in its survival in an expanding 
Universe. This finding can have other consequences. For instance, dark energy may be the ultimate 
environment for decoherence of other quantum systems in the Universe. Furthermore, the inferred 
symmetries deeply routed in the foundation of a quantum universe~\cite{houriqm} may have a prominent 
role in the formation of the fabric of spacetime and eventually direct us toward a unified model of 
gravity and quantum mechanics~\cite{qftgroup,houriqgr}.

The goal of phenomenological aspects of dark energy models summarized here has been discrimination 
among candidate models. For this purpose, we proposed a new set of parameters useful for modeling data 
to distinguish among three principle category of models. However, many other tasks remain to accomplish 
and explore. Nonlinear regime of LSS has been extensively studied using techniques such as expansion 
to nonlinear orders of perturbations~\cite{lsshigerorder}, effective field theory~\cite{lsseffectivft}, 
and peak statistics~\cite{lsspeakstat} and its bias~\cite{lssbias}. However, these works usually 
for a cosmological constant as dark energy and do not consider other possibilities. Although some 
studies investigate the effect of massive neutrinos, interaction between them components is ignored. 
This omission smears signature of interaction between dark energy and other component which is 
necessary for discriminating between models. On the other hand, parametrisation of cosmological 
observables reviewed here for discriminating is based on linear perturbations, and only applicable 
to large distant scales. However, models classified as modified gravity according to definition given 
in Sec. \ref{sec:lssdiscrim} have distinguishable imprints at intermediate and short distances. 
At these scales nonlinear effects such as bias in the peak distribution and dark matter-baryon bias 
become important and may be confused with exotic models or smear their signals. Therefore, it is 
necessary to extend the formulation to nonlinear regime. Furthermore, secondary effects such bias due 
to microlensing and what is called relativistic effects on LSS~\cite{lssrelativistic} should be 
understood and taken into account in the estimation of cosmological parameters ans search for the 
underlying dark energy model. 

\subsubsection*{\center{Acknowledgments}}
{\it I would like to thank the members of my H.D.R. jury: Niel Gehrels, Dominik Schwarz, 
Daniel Steer, Phillipe Brax, and Philippe Uzan for accepting to review my thesis - based on which 
this review is prepared - and for their support, encouragement and constructive comments.}

\end{document}